\documentclass[preprint,prb,groupedaddress]{revtex4-1}

\usepackage{amsmath}
\usepackage{graphicx}
\usepackage{multirow}

\begin{document}

\title{Nested sampling for materials: the case of hard spheres}

\author{L\'\i via B. P\'artay}
\affiliation{University Chemical Laboratory, University of Cambridge, Lensfield Road, CB2 1EW Cambridge, United Kingdom}
\email{lb415@cam.ac.uk}

\author{Albert P. Bart\'ok}
\affiliation{Engineering Laboratory, University of Cambridge, Trumpington Street, CB2 1PZ Cambridge, United Kingdom}

\author{G\'abor Cs\'anyi}
\affiliation{Engineering Laboratory, University of Cambridge, Trumpington Street, CB2 1PZ Cambridge, United Kingdom}

\date{\today}

\begin{abstract} 
The recently introduced nested sampling algorithm allows the direct and efficient calculation of the partition function of atomistic systems.  We demonstrate its applicability to condensed phase systems with periodic boundary conditions by studying the three dimensional hard sphere model. Having obtained the partition function, we show how easy it is to calculate the compressibility and the free energy as functions of the packing fraction and local order, verifying that the transition to crystallinity has a very small barrier, and that the entropic contribution of jammed states to the free energy is negligible for packing fractions above the phase transition. We quantify the previously proposed schematic phase diagram and estimate the extent of the region of jammed states. We find that within our samples, the maximally random jammed configuration is surprisingly disordered.  
\end{abstract}

\maketitle


\section{Introduction}

Nested sampling (NS) was introduced by Skilling\cite{bib:skilling,bib:skilling2} about a decade ago in the field of applied probability and inference to efficiently sample probability densities in high dimensional spaces where the regions contributing most of the probability-mass are exponentially localised. It is an annealing algorithm that  creates a sequence of probability distributions, each more concentrated than the previous one near the high-likelihood region of the parameter space (or, in the context of materials, the low energy region of configuration space) but with the crucial property that the ratio of phase space volumes that support the successive distributions is known, and is approximately constant during the annealing process.  This enables the direct evaluation of the density of states, the partition function, and all other thermodynamic observables. 

Since its original inception, nested sampling has been taken up in astrophysics \cite{multinest1} where there is a long history of association with the inference community, and it has also been applied to atomistic systems. It was shown to gain orders of magnitude in efficiency over parallel tempering (also known as replica exchange sampling) for exploring the configurations and calculating heat capacity curves of Lennard-Jones clusters and permitted the construction of phase stability diagrams.\cite{bib:our_NS_paper} Burkoff {\em et al.} used it to explore the folding landscape of proteins,\cite{Burkoff1} and Do {\em et al.} studied liquid vapour coexistence,\cite{bib:wheatley1} binary liquid mixtures\cite{bib:wheatley2} and combined it with their ``cage model'' for solids.\cite{bib:wheatley3} Recently Nielsen carried out nested sampling using a sequence of NVT trajectories.\cite{bib:NS_NVT}

We start with a brief account of the nested sampling algorithm, and to aid understanding in the materials context we loosely follow the exposition of Wheatley rather than the original formulation of Skilling. Consider the partition function $Z(V,\beta)$ of a system of $N$ particles at constant volume $V$ and inverse temperature $\beta$,
\begin{equation}
Z(V,\beta) = \frac{1}{N!}\left(\frac{2\pi m}{\beta h^2}\right)^{3N/2} \int d\mathbf{p} \, d\mathbf{q} \, e^{-\beta H(\mathbf{p},\mathbf{q},V)} ,
\label{eq:Z}
\end{equation}
where $\mathbf{q}$ and $\mathbf{p}$ are the configurational and the corresponding momentum variables, respectively, and $H$ is the Hamiltonian. Separating out the momentum-dependent part of the Hamiltonian in the usual way, we focus on the configurational partition function and using the fact that the dependence of the Hamiltonian on the coordinates $\mathbf{q}$ is solely through the potential energy $U$, we can write the partition function in terms of the density of states, $\Omega$,
\begin{eqnarray}
Z(V,\beta) &=& Z_\beta \int d\mathbf{q}\, e^{-\beta U(\mathbf{q},V)} = Z_\beta \int du\,  e^{-\beta u}\, \Omega_V(u),\label{eq:Zomega}
\\
Z_\beta &=&  \frac{1}{N!}\left(\frac{2\pi m}{\beta h^2}\right)^{3N/2},\label{eq:Zbeta}\\
\Omega_V(u) &=& \int d\mathbf{q} \, \delta(U(\mathbf{q},V)-u),\label{eq:Omega} 
\end{eqnarray}
where we have moved $V$ into the subscript of $\Omega$ to emphasise that it is a parameter rather than a variable in the present context. In order to calculate $\Omega$, let us consider its cumulative function, defined by
\begin{equation}
\gamma_V(u) = \int_{-\infty}^u du^\prime \, \Omega_V(u^\prime).
\label{eq:gam}
\end{equation}
Since the potential energy function for a given material is obviously bounded from below by the binding energy of the most stable phase, there is no problem with taking the lower limit of the integral to be negative infinity. Computing $\gamma(u)$ directly for any particular $u$  is very difficult. However, since it is the cumulative of a distribution, it is monotonic in $u$,  and therefore we are free to consider its inverse, $u(\gamma)$. Noting that all thermodynamic observables of interest depend only on the logarithm of $Z$ and therefore we are only interested in $\gamma$ and $\Omega$ up to a constant multiplicative factor, nested sampling proceeds to compute $u(\gamma)$ as follows. 
 
\begin{enumerate}
\item Select a very high potential energy value $U = u_0$, such that the structure of the corresponding subset of configuration space is little different from that of the ideal gas. Fix the corresponding value of $\gamma_0$ so that $u(\gamma_0) = u_0$, thus absorbing the arbitrary multiplicative constant in $\gamma$. Set $i=0$.
\item Obtain $K$ samples in configuration space from the {\em uniform} distribution over $\mathbf{q}$, subject to the constraint $U(\mathbf{q}) < u_i$. This is easily accomplished by Markov chain Monte Carlo and rejection sampling.
\item Compute the median of the potential energies of the $K$ samples, and let us call it $u_{i+1}$. 
The volume of the phase space covered by the samples with potential energy $U(\mathbf{q}) < u_{i+1}$ is given by the random variable $\gamma_{i+1}$ whose distribution is approximately $\mathtt{Beta}(K/2,K/2)\gamma_i$,  with a mean of $\gamma_i / 2$ whose statistical error that scales with $K^{-1/2}$. Thus we obtain the estimate $u(\langle \gamma_{i+1}\rangle) = u_{i+1}$.
\item Let $i\leftarrow i+1$ and go back to step 2. 
\end{enumerate}

Having obtained $u(\gamma)$ at a geometrically decreasing series of $\gamma$ values, we can invert it to get $\gamma(u)$ and obtain $\Omega_V(u)$ by finite difference, which can then be used to evaluate the partition function according to equation~(\ref{eq:Zomega}). The procedure hinges critically on our ability to obtain uniformly distributed samples in step 2 above. This can be aided by recycling the positions of the $K$ samplers from the previous iteration, so that only a modest number of MCMC steps are required to re-equilibrate them according to the new constraint. As was touched upon in previous works,\cite{bib:our_NS_paper,bib:NS_NVT} the number of samplers required for a given accuracy can be strongly affected by the basin-structure of the potential energy surface, because the simple MCMC rejection sampler can have difficulty in converging to the uniform distribution once the constraint results in the distribution having disconnected domains. As usual in any Monte Carlo procedure, this type of problem can be alleviated by choosing proposal distributions that are better suited to a given system. Furthermore, the risk that a disconnected domain is missed completely by the samplers is inversely proportional to the size of the domain (and also to the number of samplers of course), and therefore large domains are less likely to be missed. This tradeoff between the number of samples generated and the risk of missing relevant configurations exists in all stochastic simulation methods, and it is no better or worse in nested sampling.

It is helpful to compare the nested sampling with simulated annealing, which also creates a series of overlapping distributions, each one proportional to the Boltzmann measure with ever decreasing temperature parameters. The major difference is that the average phase space volume ratio of the successive distributions, which is needed to calculate the partition function, is directly available (indeed set by the user) in nested sampling, whereas it would need to be computed in simulated annealing, and if the successive distributions have little overlap, the estimate of this ratio would have an enormous error. This problem would be particularly severe near a first order phase transition, where a small change in the temperature corresponds to a very large change in the distribution, and therefore the overlap between the successive distributions across the phase transition vanishes in the thermodynamic limit. For this reason, the partition function is not considered accessible by simulated annealing. In contrast, nested sampling will maintain the distribution overlap by design, even through a phase transition.

Practitioners (and Skilling originally) often replace the median in step 3 with the $K/(K+1)$ fraction of the potential energy distribution of the $K$ samples, which does not change the algorithm substantially and results in a finer sampling of the $u(\gamma)$ function. In this case, the next energy level $u_{i+1}$ is simply given by the potential energy of the sample with the highest value in the sample set, and the corresponding phase space volume is $\gamma_{i+1} = \mathtt{Beta}(K,1) \gamma_i$ (note that $\mathtt{Beta}(K,1)$ is equivalent to $\mathtt{Uniform}(0,1)^{1/K}$), thus $\langle \gamma_{i+1} \rangle= \langle \gamma_i \rangle K/(K+1)$. 

The errors of any quantity derived from the partition function are most easily computed by generating a large dataset using ``bootstrapping'', i.e. resampling the phase space volume ratios, $\gamma$, according to their theoretical distribution.\cite{bib:skilling_convergence} The mean and standard error of the thermodynamic quantities are then given by the distribution of the sample mean.

The nested sampling iterations can in principle be continued indefinitely, obtaining ever lower energy samples, however due to the bounded nature of the potential energy for atomistic systems, configurations whose energy is very close to the lower bound will only become relevant for computing thermodynamic observables at very low temperatures. Therefore a natural and very practical criterion for terminating the iterations is to aim for the convergence of desired observables for all temperature values above a preset value. 

The purpose of the present work is to demonstrate the nested sampling technique for condensed phase systems. This is different from previous applications because the ordered/disordered phase transition is much sharper than it was in the case of isolated clusters, and therefore is particularly challenging to the annealing algorithm, while still having a macroscopic number of local minima in the configuration space. The hard sphere system presents these essential features of a condensed phase material. Much is known about the hard sphere system in the literature, and indeed this contributed to our choice to use it as an exemplar. The significance of present work lies in the {\em power and simplicity} of nested sampling algorithm, in that it yields the partition function (and thus all desirable thermodynamics observables) in a straightforward manner, thus unifying the methodology for calculating various quantities. This makes it eminently suitable for high throughput calculations in the future. 

\subsection{The hard sphere system}
One of the first material systems ever studied using computer simulation was the hard sphere model and its  simpler two-dimensional version consisting of hard disks.\cite{bib:HS_Rosenbluth, bib:HS_Wood, bib:HS_Wainwright} Since then hard spheres have been widely used as model particles not just in the field of statistical mechanics,\cite{bib:HS_phase_transition} but also in biology,\cite{HS_bio, HS_bio2} engineering and material science,\cite{HS_eng,Composit_model} and consequently extensively studied by analytical means,\cite{HS_anal} computer simulations,\cite{bib:HS_interface,bib:HS_stacking_faults, bib:HS_coexistence_Noya,bib:HS_crystallization,bib:HS_crystallization2,bib:HS_noglass,EoS_Alder,EoS_Santos,EoS_Wang,EoS_WC1,Bolhuis_fcc-hcp} and also  experimental methods using colloids.\cite{HS_exp_PMMA,bib:HS_crystallization_exp,bib:HS_crystallization_exp2,bib:HS_crystallization_exp3,HS_exp_spaceshuttle,Zhu_spaceshuttle,Lekker_diff}

The hard sphere system undergoes a first order solid-fluid  phase transition\cite{bib:HS_phase_transition} as the packing fraction $\phi$ is increased. The fluid phase is stable below the freezing point, $\phi \approx 0.494$, beyond which  fluid and solid  coexist up to the melting point at $\phi \approx 0.545 $.\cite{bib:HS_phase_transition, bib:HS_coexistence_Noya} The densest possible packing fraction which can be achieved with identical hard spheres is $\pi / 3 \sqrt 2 \approx 0.74048$, corresponding to the fcc lattice and its stacking variants --- this was known as Kepler's conjecture and has recently been proved.\cite{bib:Keplers_proof} However, metastable solid structures exist (both experimentally and in simulations) where the particles are disordered at relatively high density, in the so-called \emph{random close packed} (RCP) state. Due to the proposed similarity of this disordered phase to glassy phases in more complex systems, these structures are widely studied, even though there is no glass transition in one-component monodisperse hard sphere systems.\cite{bib:HS_noglass_gr,bib:HS_noglass,bib:HS_crystallization} It is known from experimental studies\cite{bib:HS_crystallization_exp,bib:HS_crystallization_exp2,bib:HS_crystallization_exp3} and computer simulations \cite{bib:HS_crystallization, bib:HS_crystallization2} that the disordered phase is thermodynamically unstable with respect to the ordered crystal.

\begin{figure}[bt]
\begin{center}
\includegraphics[width=8cm]{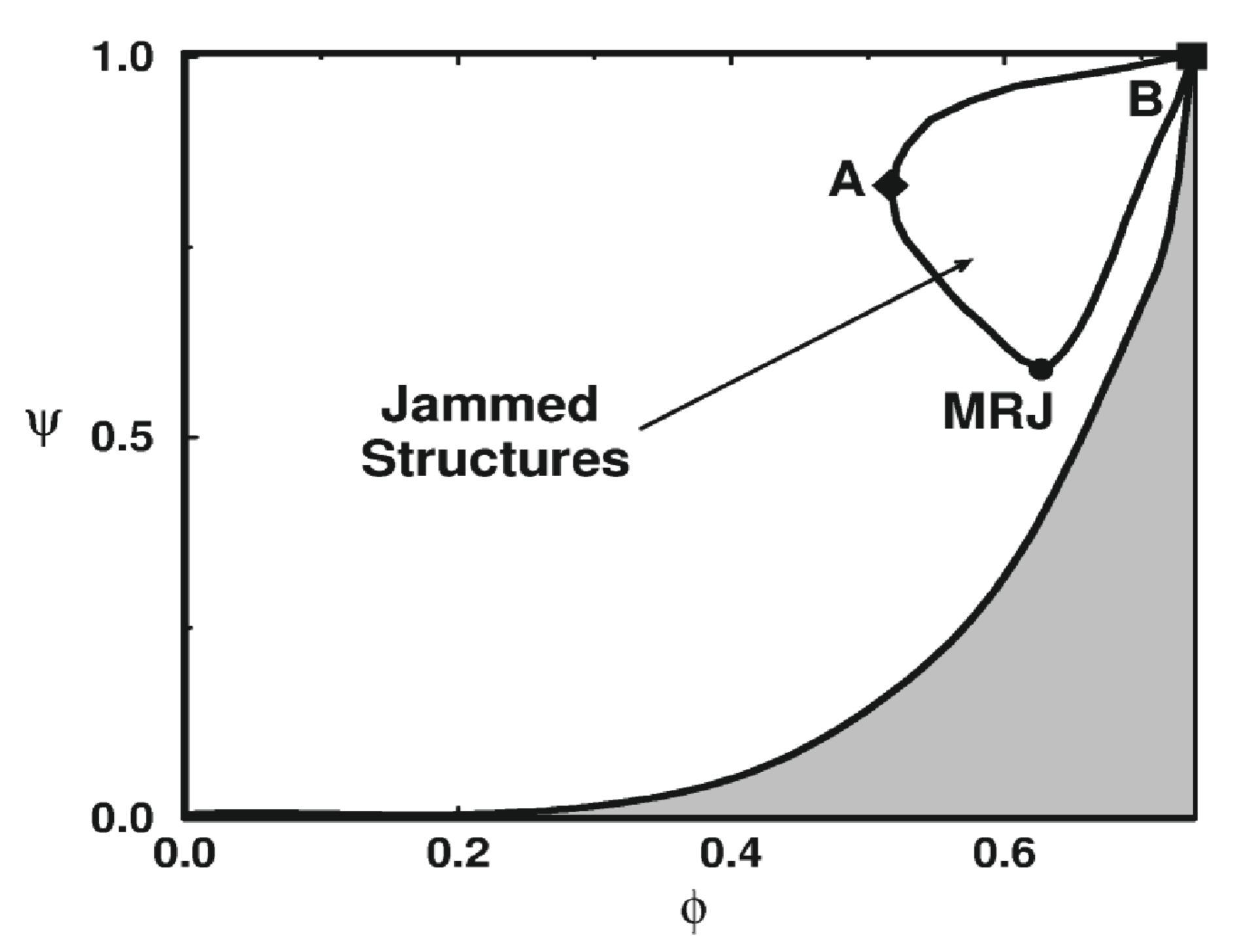}
\end{center}
\caption {Schematic plot of an arbitrary order parameter, $\Psi$, of hard sphere configurations as a function of the packing fraction, $\phi$. The grey area represents the inaccessible region, and the black squares show the maximally random jammed state (MRJ), the jammed state with the lowest possible packing fraction (A) and the perfect crystalline structure (B). (Reprint of Fig 1. of Ref.~[\citep{bib:debenedetti_RCP}], with the permission of the authors. Copyright 2000 by The American Physical Society.)}
\label{fig:Order_packingfraction}
\end{figure}

Although the term ``random close packed'' is widely used in the literature, Torquato and his co-workers pointed out that the RCP state is very ill defined,\cite{bib:debenedetti_RCP} and the corresponding packing fraction that is reported varies in the range of $\phi \approx 0.60-0.68$ depending on the protocol used to produce the configurations.\cite{bib:RCP_exp, bib:RCP_exp2, bib:RCP_calc_densification,bib:RCP_calc_MC,bib:RCP_calc_dropandroll}
The problem lies in the fact that it is not well defined how random the ``random'', or how close packed the ``close packed'' really is. They suggest that the randomness should be quantified by an order parameter, and the term ``close packed'' should be replaced by the term {\em jammed}. A system is jammed, if all the particles are jammed in the system, and a particle is jammed if it cannot be translated while the rest of the system is kept fixed. The various states can be sketched on a two dimensional diagram using the packing fraction $\phi$ and an (as yet undefined, ``ideal'') order parameter $\Psi$ as axes, as shown in Figure~\ref{fig:Order_packingfraction}, reproduced from \cite{bib:debenedetti_RCP}. The bottom left corner represents the completely disordered spheres of zero size, while the top right corner $(\phi \approx 0.74048)$ corresponds to the perfectly ordered and close packed fcc crystal. The grey area represents the inaccessible region where the hard spheres would overlap and all the jammed configurations are located within the loop. Two specific points of the loop boundary can be identified: the point marked with A corresponds to the jammed structure with the lowest possible packing fraction, and the maximally random jammed (MRJ) state is the jammed structure having the lowest possible order parameter. More recently, Torquato et al. refined the definition of jammed states and distinguished three sub-categories (locally, collectively and strictly jammed), \cite{bib:torquato_jamming} and discussed the above figure in more details.\cite{bib:torquato_RCP}

In what follows, nested sampling enables us to quantitatively substantiate the sketch of Figure~\ref{fig:Order_packingfraction}, estimate the parameters of the extremal points of the jammed phase and also calculate the free energies of the various states as a function of the packing fraction. 


\section{Thermodynamics of hard spheres}

We now  briefly recall the fundamental thermodynamics of hard spheres. The interesting intensive variable is the pressure $p$, rather than the inverse temperature $\beta$, and so we consider the 
 isotherm-isobar partition function, $\Delta$, that is given by

\begin{equation}\label{eq:partition_function1}
\Delta(\beta, p, N) = \frac {1} {h^{3N}N!} \, \beta p \int \exp[-\beta(H(\mathbf{p,q},V) + pV] \, d\mathbf{p} \, d\mathbf{q} \, dV.
\end{equation}
After again taking out the momentum dependent part, $Z_p$, we change to fractional coordinates $\mathbf{s}$:
\begin{equation}\label{eq:partition_function3}
\Delta(\beta, p, N) = \beta p \, Z_p(\beta) \int_{(0,1)^{3N}} d\mathbf{s} \, \int_0^\infty dV \, V^N  \exp[-\beta(U(\mathbf{s},V) + pV].
\end{equation}
For hard spheres of a given radius, the potential energy is either zero or infinity that, for a given set of fractional coordinates, depends on the volume,
\begin{equation}\label{eq:hs_case}
U(\mathbf{s},V) = \left\{ \begin{array}{rl}
 0 &\mbox{ if $V \ge V_c(\mathbf{s})$} \\
 \infty &\mbox{ if $V < V_c(\mathbf{s})$}
       \end{array} \right.,
\end{equation}
where $V_c(\mathbf{s})$ is the function that gives the lowest permissible volume for the configuration represented by the fractional coordinates $\mathbf{s}$. Substituting into eq.~\ref{eq:partition_function3},
\begin{equation}\label{eq:partition_function3b}
\Delta(\beta, p, N) = \beta p \, Z_p(\beta) \int_{(0,1)^{3N}} d\mathbf{s} \, \int_{V_c(\mathbf{s})}^\infty dV \, V^N  \exp[-\beta pV],
\end{equation}
and now the last integral can be transformed to a standard exponential integral, 
\begin{equation}\label{eq:partition_function3c}
\int_{V_c(\mathbf{s})}^\infty dV \, V^N  \exp[-\beta pV] = V_c(\mathbf{s})^{N+1} \, E_{-N} (\beta V_c(\mathbf{s}) p),
\end{equation}
\begin{equation}
E_N(x) = \int_1^\infty dt \frac {e^{-xt} } {t^N}
\end{equation}

The partition function thus becomes
\begin{align}
\Delta(\beta, p, N) &= \beta p \, Z_p(\beta) \int_{(0,1)^{3N}} d\mathbf{s} \, V_c(\mathbf{s})^{N+1} \, E_{-N} (\beta V_c(\mathbf{s}) p) \\
&= \frac {1} {\beta^N p^N} \, Z_p(\beta) \int_{(0,1)^{3N}} d\mathbf{s} \, \Gamma(N+1,\beta V_c(\mathbf{s}) p),\label{eq:posint}
\end{align}
where $\Gamma$ is the upper incomplete gamma function,
\begin{equation}
\Gamma(N,x)=\int_x^\infty t^{N-1} e^{-t} \,\mathrm{d}t = x^N E_{1-N}(x)
\textrm{.}
\end{equation}

For completeness, recall that given the partition function, thermodynamic observables can be computed easily. For example, the Gibbs function is given by
\begin{equation}
G(T,p,N)=-k T \ln \Delta,
\end{equation}
and so the expected volume and the compressibility are,
\begin{eqnarray}
V&=&\frac{\partial G}{\partial p}=-k T \frac{1}{\Delta} \frac{\partial \Delta}{\partial p}\\
\kappa &=& - \frac {1} {V} \frac {\partial V} {\partial p}. \label{eq:compressibility}
\end{eqnarray}

 We compute the final partition function by combining the density of states samples, $\Omega_V(u)$, and associated weights, $\gamma(u)$, (see eqs~\ref{eq:gam} and~\ref{eq:Zomega}) from a number of independent simulations.
 

\section{Nested Sampling of hard spheres}

When applying nested sampling to the hard sphere problem, the role of the potential energy $U$ in the algorithm in the Introduction is taken  by $V_c(\mathbf{s})$, and the density of states that is obtained is used to calculate the partition function according to equation~(\ref{eq:posint}). Notice that the temperature $\beta$ and pressure $p$ occur only as a product in this equation, and therefore a single nested sampling annealing run enables us to calculate observables at all pressures and temperatures. 

The size of the sample set, $K$, and the number of attempted moves, $M$, during the MCMC rejection sampling in step 2 of the algorithm are convergence parameters. In the present work, $K$ was between 500 and 1500, and $M$ was between 80 and 320. The proposal distribution for the MCMC sampler was a Gaussian in the position of a single particle. The variance of the Gaussian was periodically adjusted to maintain a constant acceptance rate. In order to quantify finite size effects, systems of 32, 64 and 128 hard spheres were used. For each system size we performed four independent calculations starting from different initial conditions. The total number of MC steps needed to converge the partition function and to reliably reach the crystalline structure with the highest possible packing fraction are shown in Table~\ref{table:convergence}. We emphasise again, that after performing the above number of MC steps, arbitrary thermodynamic variables can be calculated at any given packing fraction, without further computational cost. In comparison, Noya et al. \cite{bib:HS_coexistence_Noya} performed $4 \times10^9-5 \times10^{10}$ MC steps to converge a solid/liquid coexistence simulation at a \emph{single} value of the packing fraction, using a unit cell of 5184 particles. 

\begin{table}
\caption{Parameters used to produce the converged nested sampling simulations for different number of particles, $N$.  $K$ is the number of walkers and $M$ is the number of random walk steps in each iteration. The total number of MC steps is equivalent to the number of evaluations of the $V_c(\mathbf{s})$ function during the whole simulation.}
\begin{tabular}{cccc}
& \\ [-1.5ex]
\hline\hline
$N$ &$K$& $M$ &MC steps \\
\hline
 32        & 500   & 80& $5.3 \times 10^8$    \\  %
 64      & 1000 & 80  & $3.9 \times 10^{9}$  \\  %
128      & 1500 & 320 & $8.0 \times 10^{10}$\\  %
\hline\hline
\end{tabular}
\label{table:convergence}
\end{table}


\section{Compressibility, Equation of State and Free energy}

\begin{figure}[btp]
\begin{center}
\includegraphics[width=9.5cm]{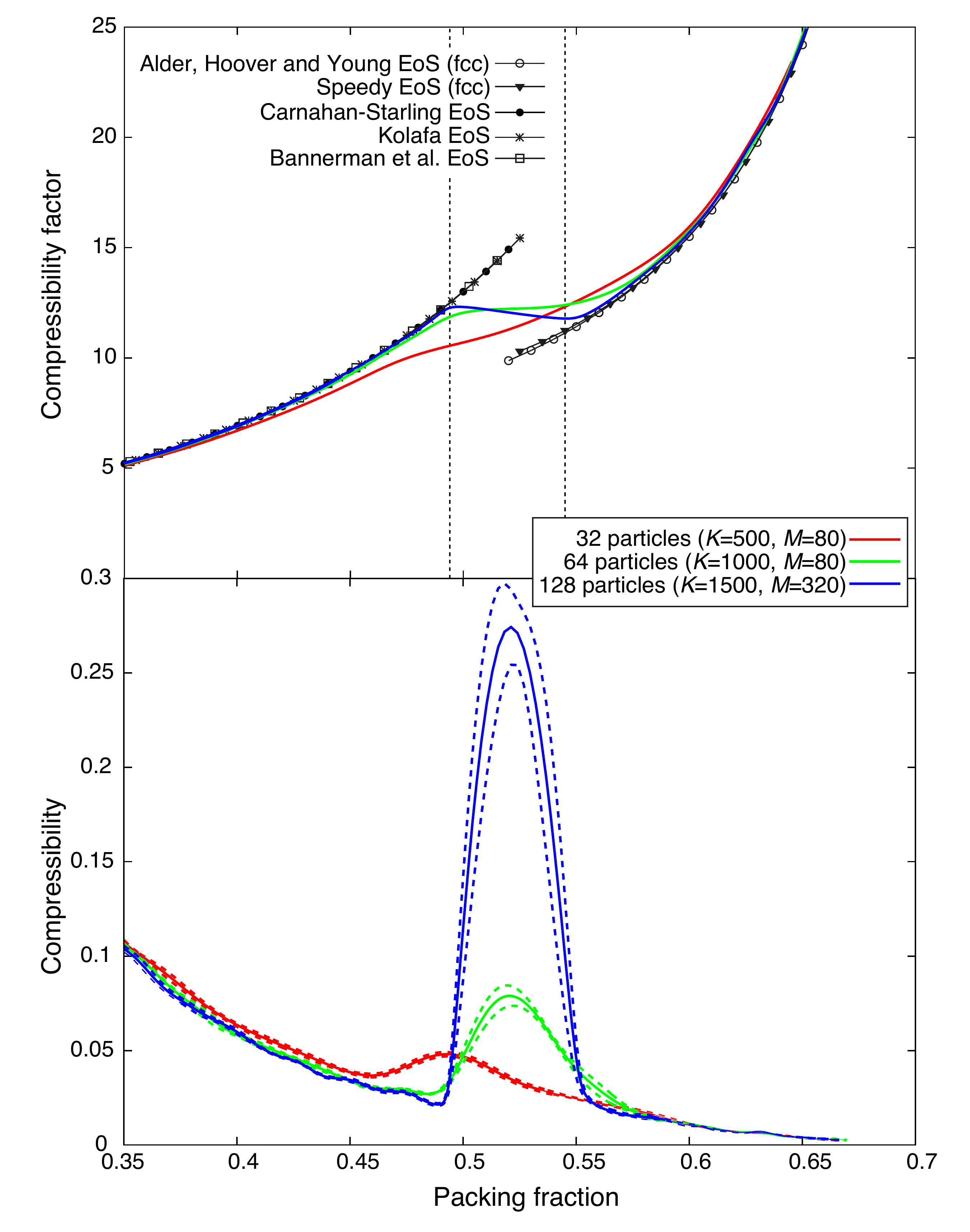}
\end{center}
\caption {Compressibility factor (top panel) and the compressibility itself (in units of $\beta \sigma^{3}$, bottom panel) as a function of packing fraction from our nested sampling runs, for 32 (red), 64 (green), 128 (blue) hard sphere particles. Five well known equations of state for hard spheres are shown for comparison with black curves and different symbols. The vertical dashed lines indicate the packing fraction values of freezing and melting. The error of the compressibility factor is on the order of the line thickness, and for the compressibility the standard errors are shown by coloured dashed lines.}
\label{fig:HS_compressibility}
\end{figure}

Given the partition function, it is now easy to evaluate thermodynamic observables, such as the compressibility, $\kappa$, and the compressibility factor, $z = {pV} /{Nk_BT}$ (i.e. the equation of state, EoS) as a function of packing fraction, which are shown in Figure~\ref{fig:HS_compressibility}. There is no exact theoretical solution for the EoS of the hard sphere system, and a huge variety of fits can be found in the literature, separately for the fluid and solid phases.\cite{EoS_Wang,bib:Mulera_HSEoS_sum} Nested sampling, on the other hand, gives the equation of state naturally for the whole range of packing densities. The liquid branch is compared to the simple but successful Carnahan-Stirling EoS~\cite{EoS_CS}, to the EoS proposed by Kolafa~\cite{EoS_Kolafa}, and to one of the most recently published closed-virial EoS, the work of Bannerman et al.~\cite{EoS_WC1}. The solid branch is compared to the EoS of Speedy\cite{EoS_Speedy} and Alder et al.\cite{EoS_Alder}, both developed to predict the behaviour of the fcc crystal. The compressibility shows the expected peak at $\phi = 0.49-0.55$, in the coexistence region.\cite{bib:HS_phase_transition,bib:HS_coexistence_Noya} Both the compressibility and the compressibility factor clearly have a significant finite size error for 32 particles, but it is remarkable that the finite size error of the nested sampling simulation is already very small for $N=64$, in contrast to what is expected in the case of explicit coexistence simulations. 

\begin{figure}[bthp]
\begin{center}
\includegraphics[width=9.5cm]{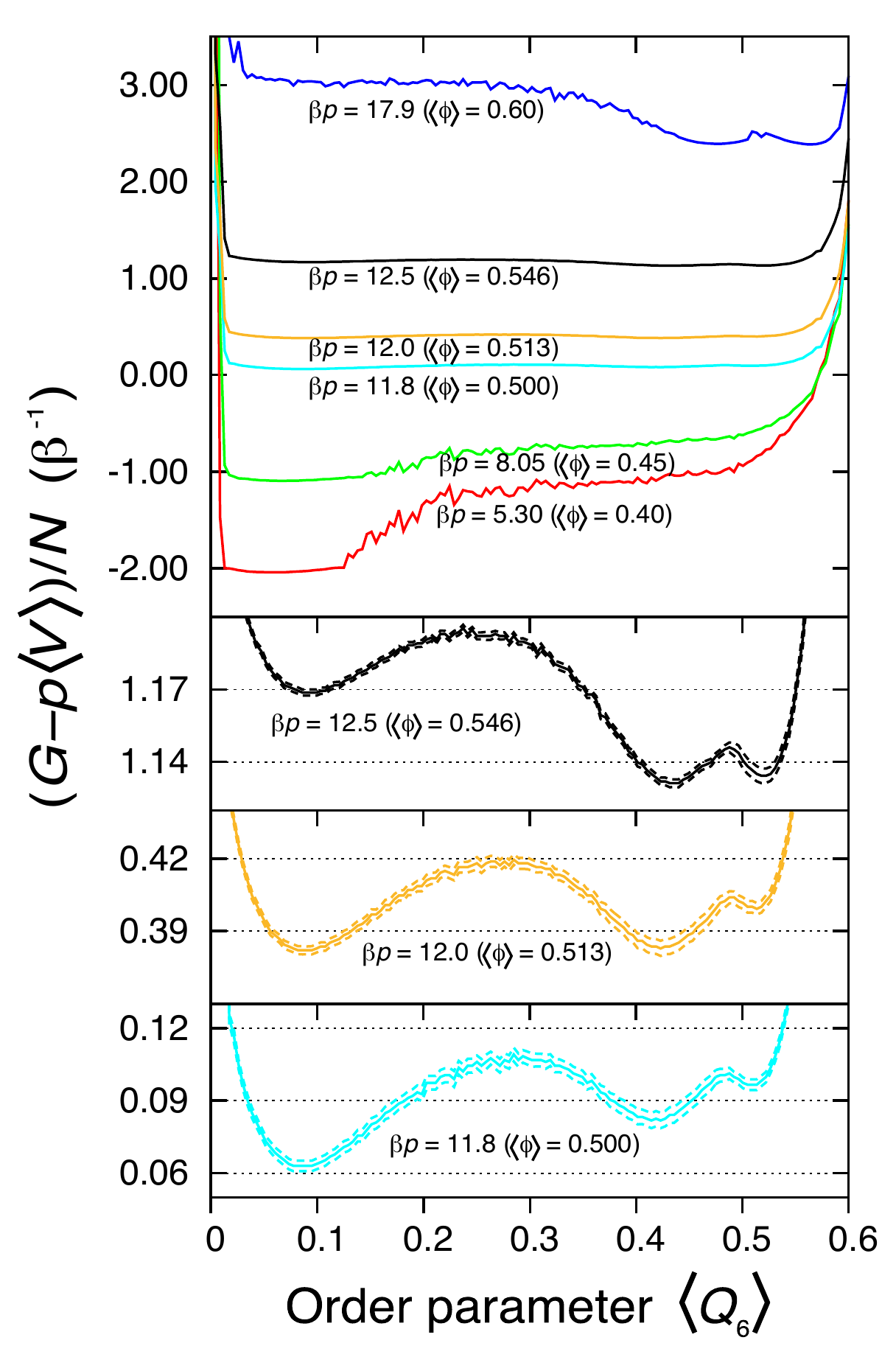}
\end{center}
\caption {Gibbs free energy per particle of a periodic lattice of hard spheres with 128 particles in the unit cell as a function of the $Q_6$ order parameter at different  pressure values ($\beta p$ in $\sigma^{-3}$ units) calculated by eq.~\ref{eq:G_free_energy}. In order to show the curves for different pressure values on the same plot, $-p\langle V\rangle$ was added to the free energy. The corresponding packing fraction ($\phi$) values are also indicated. The bottom three panels show the free energy curves in a larger scale around the phase transition, and the error is shown by dashed lines.}
\label{fig:HS128_G_free_energy}
\end{figure}

Access to the partition function enables the calculation of absolute free energies as a post-processing step after the nested sampling simulation is complete, therefore we are able to compute the Gibbs free energy of the hard sphere system. The natural variable of the Gibbs free energy is the pressure, and to shed light on the phase transition, we consider  the well known bond order parameter $Q_6$\cite{bib:Q6parameter} as a further thermodynamic variable. Its value is zero for completely disordered systems and increases with degrees of crystallisation. The $Q_6$ value differs for the perfect fcc and hcp crystals and their stacking variants, $Q_6(\mbox{fcc}) = 0.57452$ and $Q_6(\mbox{hcp}) = 0.48476$. Thus the Gibbs free energy becomes, up to an additive constant, 
\begin{equation}\label{eq:G_free_energy}
G(p,Q_6) = -\beta^{-1} \ln (\Delta(\beta, p,Q_6)) \approx -\beta^{-1} \ln \left [ \int d\mathbf{s}\,
\Gamma(N+1, \beta p V_c(\mathbf{s}) \delta(Q_6(\mathbf{s}) - Q_6)) \right ].
\end{equation}  
The calculated Gibbs free energies are shown in Figure~\ref{fig:HS128_G_free_energy} as a function of $Q_6$ for a series of fixed pressure values (with the corresponding average packing fraction in parentheses) in case of the 128 particle system. The free energy barrier corresponding to the phase transition at $p \approx 12.0, \phi \approx 0.513$ is seen to be small, $0.04\,\beta^{-1}$ per particle. 

At low pressures (below $p \approx 8.05, \phi \approx 0.45$) there is only one free energy minimum present, corresponding to the fluid phase. As the pressure and packing fraction increases, two distinct free energy minima appear, corresponding to the hcp and fcc stacking orders. Although the two structures have the same density at maximum packing, their thermodynamic properties differ due to the difference in their second and subsequent shell neighbours.\cite{Pronk_elastic_diff} The relative stability of the two stacking variants had remained a long-standing question, until the fcc structure was found to be more stable than hcp,\cite{Woodcock_hcpfcc} however, the Helmholtz free energy difference is only $9 \times 10^{-4} \,\beta^{-1}$ per particle at the melting point, $\phi = 0.545$.\cite{Bolhuis_fcc-hcp} At this packing fraction we found that the free energy difference is within the error of our calculations (see second panel on Fig.~\ref{fig:HS128_G_free_energy}), and the difference is the same order of magnitude as the literature value.  
More interestingly, at lower packing fractions, $\phi \approx 0.513$ and $\phi \approx 0.50$, we found larger free energy differences, and in favour of the hcp structure. Although the difference is still small, it might explain the experimental observation that hard-sphere colloids rarely crystallise directly into the fcc structure, but initially form a randomly stacked hexagonal close-packed (rhcp) crystal \cite{Zhu_spaceshuttle,Lekker_diff}, which then slowly transforms to the stable fcc phase.\cite{bib:HS_stacking_faults,PhysRevLett.88.015501}

\section{Order parameter vs packing fraction}

\begin{figure}[bthp]
\includegraphics[width=5.5cm]{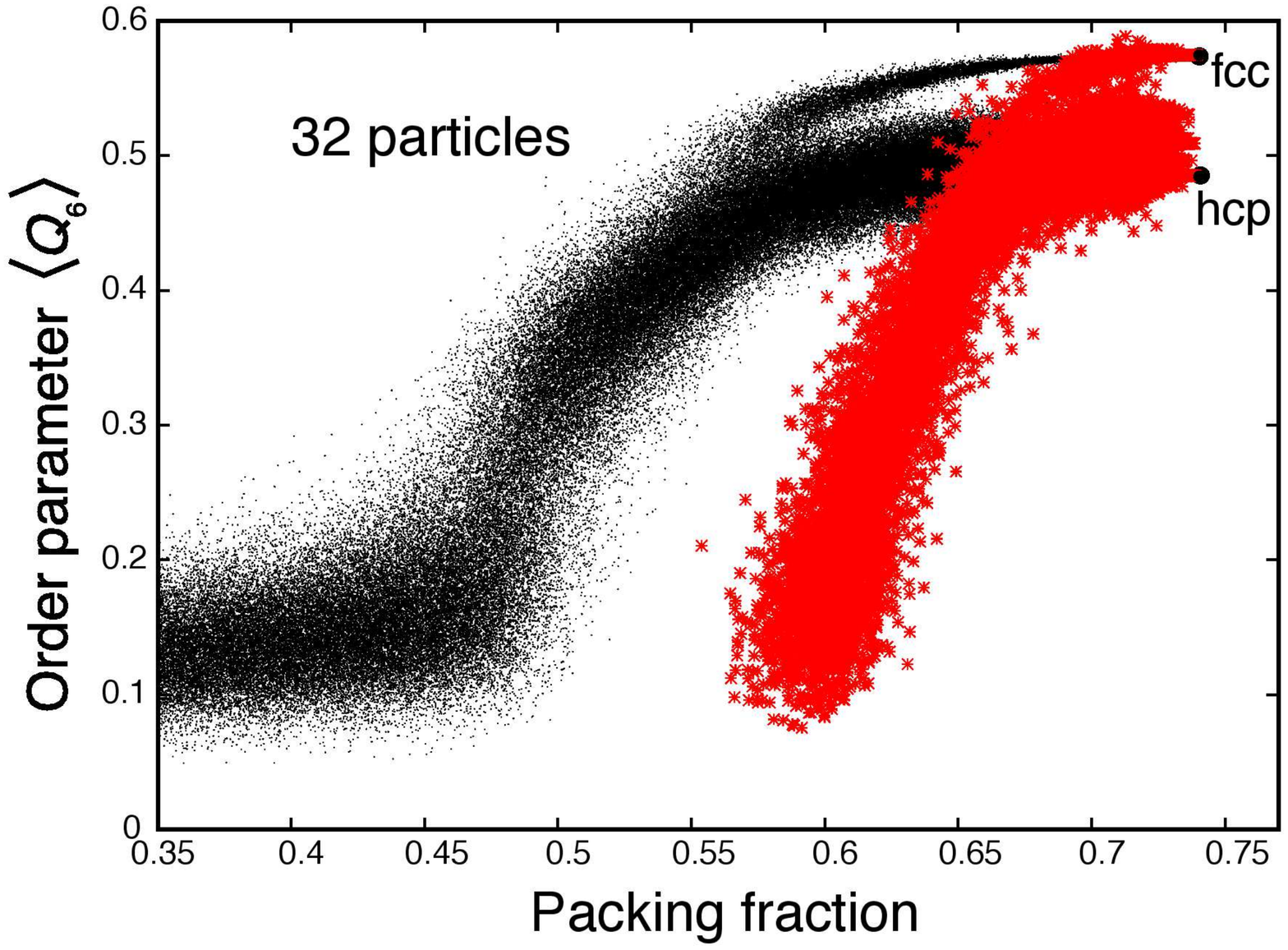}
\includegraphics[width=5.5cm]{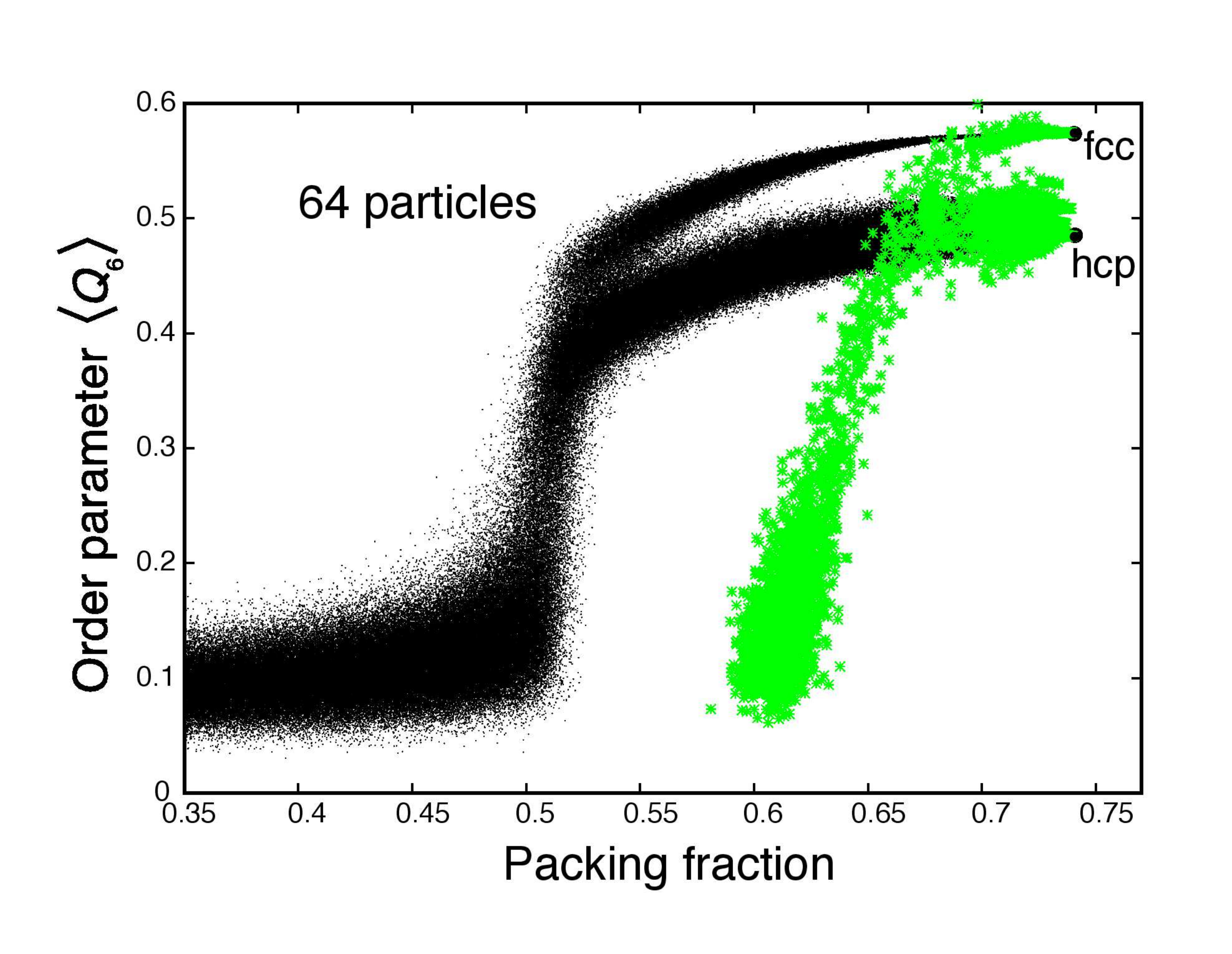}
\includegraphics[width=5.5cm]{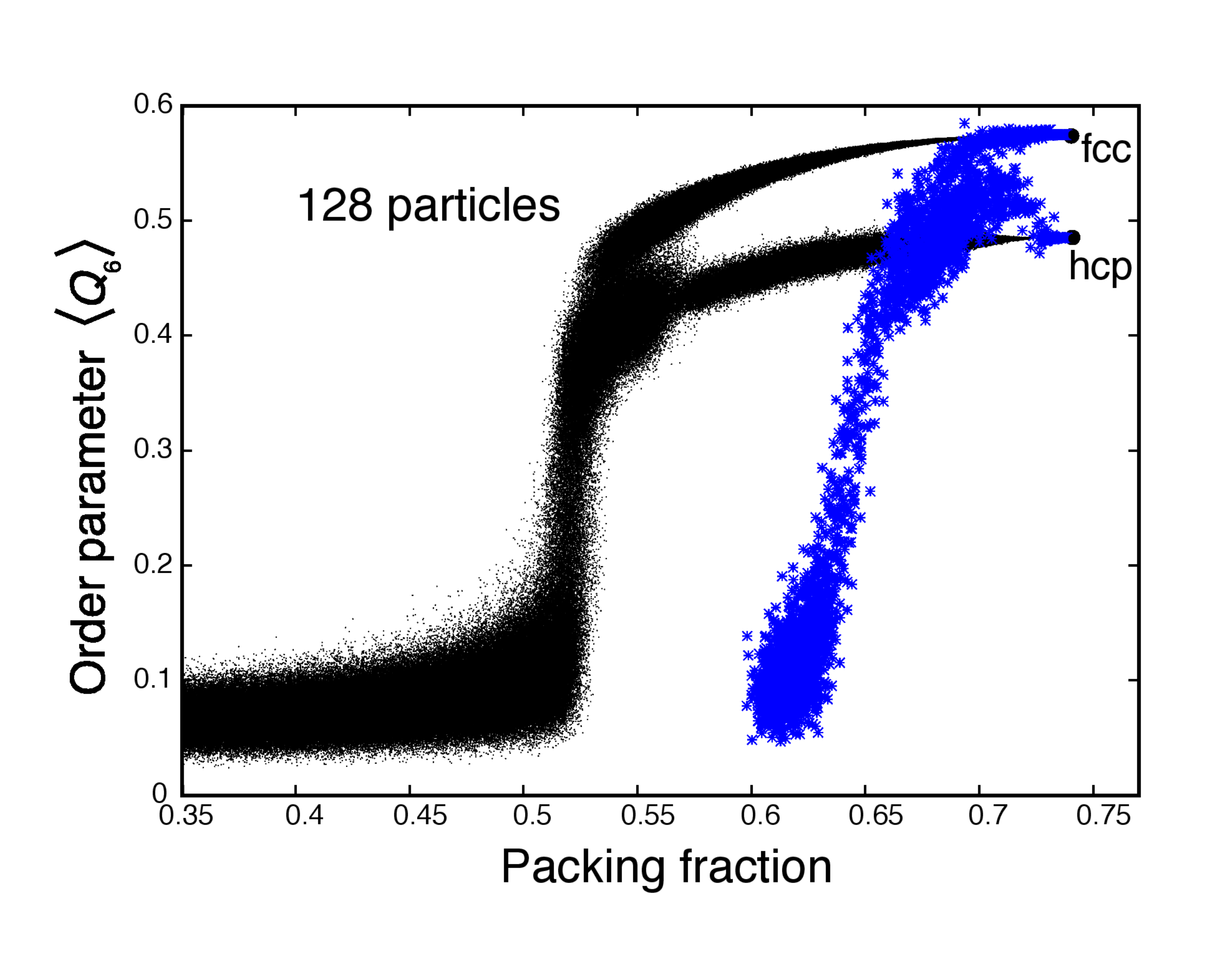}
\includegraphics[width=5.5cm]{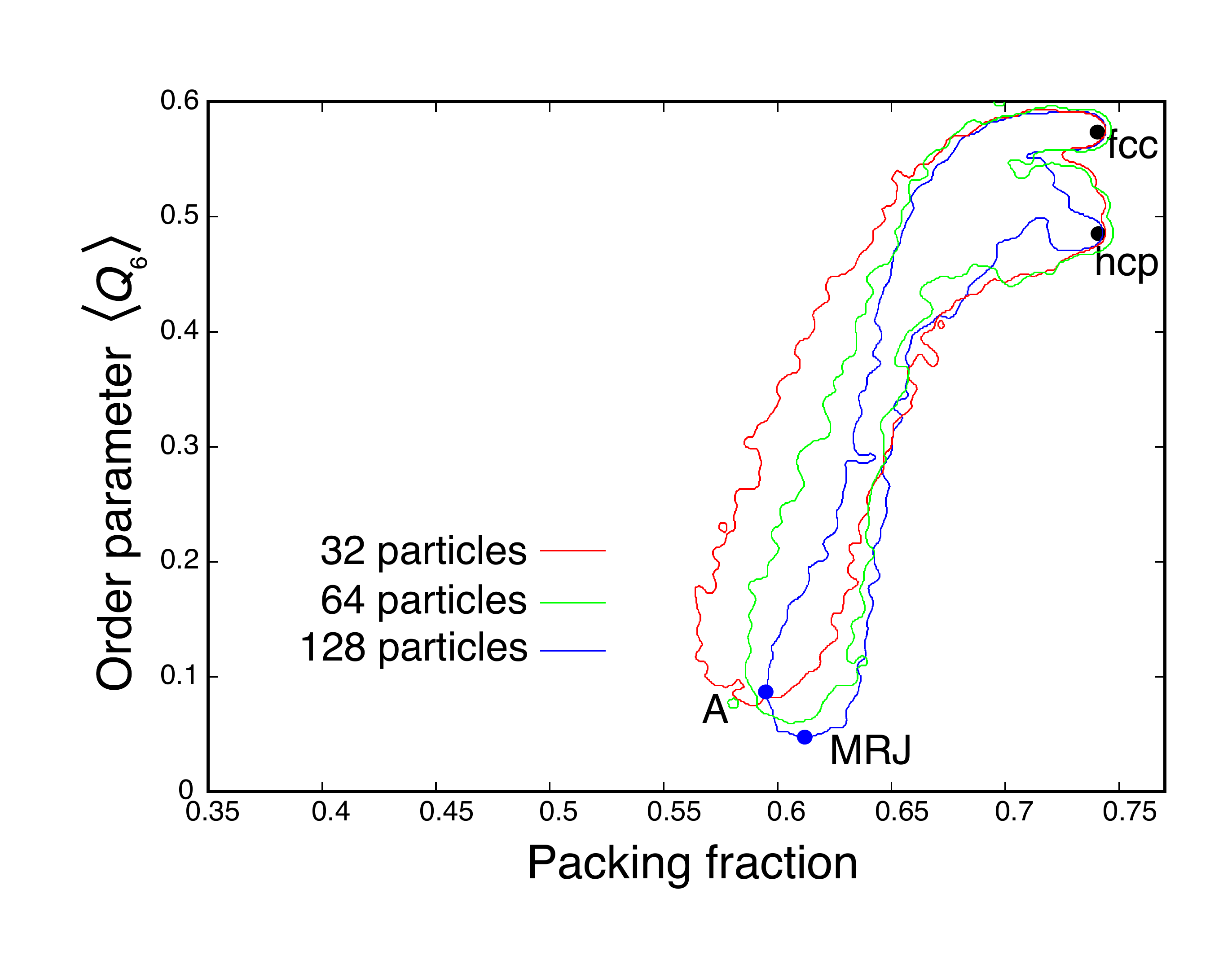}
\caption {Order parameter $Q_6$ as a function of the packing fraction for $N=32, 64 \textrm{ and } 128$ hard spheres. The black dots represent the configurations generated during the nested sampling runs and the coloured symbols represent the quenched jammed configurations (see text). The last panel shows the area occupied by the jammed states of 32 (red), 64 (green), 128 (blue) using a contour line after a little gaussian smearing. Perfect fcc and hcp structures are indicated by filled black circles, and the points MRJ (maximally random jammed state found) and A (jammed state with lowest possible packing fraction found) are also shown on the last panel for the 128 particle system.}
\label{fig:HS_orderparameter}
\end{figure}

To quantitatively substantiate the diagram of Figure~\ref{fig:Order_packingfraction}, we again use the bond order parameter $Q_6$ as the order parameter for the vertical axis. 
The calculated $Q_6$ parameters of the nested sampling configurations as a function of their packing fraction are shown with black dots in Figure~\ref{fig:HS_orderparameter} for all system sizes. The phase transition corresponding to the compressibility peak near $\phi \approx 0.49$ is revealed as a sharp increase in crystallinity. 

Nested sampling allocates samples to various regions of phase space in proportion to the region's phase space volume. Thus white areas in Figure~\ref{fig:HS_orderparameter} (i.e. regions with no samples in our simulation) correspond to regions with very small phase space volume (approximately $< 1/K$ for a given packing fraction). Accordingly, at low packing fraction, the crystalline states while clearly accessible, occupy a negligible phase space volume. Conversely, at high packing fraction, the disordered states have negligible volume, which manifests itself in the well known result that the disordered phases are thermodynamically unstable here. The samples with the highest possible  packing fraction fall into distinct groups with different order parameter values, because the nested sampling simulation found both the fcc and the hcp phases. However, it is interesting to see that the branches leading to either stacking variants of the crystal separate at relatively small $\phi$ values, already around the melting point.

To distinguish the region of jammed states from the totally inaccessible region of high packing fraction and low order at the right hand side of the plot, we chose 10000 points from the nested sampling configurations in the packing fraction range of $0.40-0.73$ and rapidly quenched them by increasing the particle diameter (and hence the packing fraction) with minimal change in the location of particle centres until the entire system became jammed. Further details on this procedure are given in the Appendix.

The final jammed configurations are plotted by coloured stars in Figure~\ref{fig:HS_orderparameter}. The size and shape of the region they occupy is converges with increasing particle number to the region of RCP states. It can be seen from these figures that jammed configurations exist in states corresponding to a wide variety of order and packing fraction. There are jammed states with packing fraction smaller than 0.6, and many others are as disordered as the fluid phase, or almost perfect crystals containing different defects. 

To further characterise the structure of the extremal configurations ``A'' and ``MRJ'' of the jammed region found in our simulations, we show the fractions of particles with varying numbers of neighbours in Table~\ref{tab:neighs}. The distribution of the number of neighbours peaks near 5 and 6 neighbours. To put the neighbour count in context we note that in the closed packed crystalline phases particles have 12 neighbours each, and 8 in the bcc phase. 

\begin{table}
\caption{Proportion of particles with given number of neighbours in the extremal configurations ``A'' and ``MRJ'' of Figure~\ref{fig:HS_orderparameter}. The number of neighbours was counted with a tolerance of 1\% on the inter-particle distance. Proportions less than 5\% are likely unreliable and thus are omitted, and percentage values are rounded.}
\begin{tabular}{c|ccc|ccccc}
\hline\hline
&&&Mean number&\multicolumn{5}{c}{Number of neighbours}\\
Configuration & $\phi$ & $Q_6$ & of neighbours & 4 & 5 & 6 & 7 & 8 \\
\hline
MRJ &0.612&0.046& 5.3 &14\% & 21\% & 22\% & 23\% & 9\% \\
A &0.598&0.083& 5.6 & 13\% & 32\% & 23\% & 23\% & - \\
\hline\hline
\end{tabular}
\label{tab:neighs}
\end{table}


\section{Conclusion}

We have demonstrated the nested sampling algorithm in the condensed phase by applying to the three dimensional periodic hard sphere system, and studied the solid/liquid phase transition. Remarkably, the observed finite size effect is small so that  both the location of the melting and freezing points and the compressibility and equation of state as a function of packing fraction are reasonably well converged for 128 particles. Coexistence simulations using a similar number of Monte Carlo moves and  $\sim 50$ times more particles obtained just the location of the transition\cite{bib:HS_coexistence_Noya}. 

Using the samples collected, we were able to plot the Gibbs free energy as a function of order parameter and packing fraction, showing the phase transition explicitly and quantitatively, demonstrating that the barrier for transition to crystallinity is small, and that above the phase transition the entropic contribution of the jammed states to the free energy is negligible. The global unbiased nature of this sampling scheme allowed us to study the phase diagram using the packing fraction and the $Q_6$ order parameter as thermodynamic variables, identifying the extent of the jammed phase. The ``maximally random jammed'' state we found, i.e. the jammed state with the lowest $Q_6$ order parameter, and the least dense jammed state, are both surprisingly disordered, having $Q_6$ values of less than 0.1. While we do not claim that there are no remaining finite size effects in the 128 particle system for any observable, the extent of the ``forbidden region'' in the phase diagram of Figure 4 shows little change in going from 32 to 128 particles. 

The ease and simplicity of nested sampling combined with its power to calculate the partition function directly and all thermodynamic observables that follow make it an excellent tool to use in computational statistical mechanics of particle systems in the condensed phase. 

\begin{acknowledgements}
The authors are grateful to Daan Frenkel for early discussions on the statistical mechanics of hard spheres. LBP thanks St. Catharine's College, Cambridge. APB thanks Magdalene College, Cambridge. 
\end{acknowledgements}

\bibliography{NSHS_reference}

\begin{thebibliography}{55}%
\makeatletter
\providecommand \@ifxundefined [1]{%
 \@ifx{#1\undefined}
}%
\providecommand \@ifnum [1]{%
 \ifnum #1\expandafter \@firstoftwo
 \else \expandafter \@secondoftwo
 \fi
}%
\providecommand \@ifx [1]{%
 \ifx #1\expandafter \@firstoftwo
 \else \expandafter \@secondoftwo
 \fi
}%
\providecommand \natexlab [1]{#1}%
\providecommand \enquote  [1]{``#1''}%
\providecommand \bibnamefont  [1]{#1}%
\providecommand \bibfnamefont [1]{#1}%
\providecommand \citenamefont [1]{#1}%
\providecommand \href@noop [0]{\@secondoftwo}%
\providecommand \href [0]{\begingroup \@sanitize@url \@href}%
\providecommand \@href[1]{\@@startlink{#1}\@@href}%
\providecommand \@@href[1]{\endgroup#1\@@endlink}%
\providecommand \@sanitize@url [0]{\catcode `\\12\catcode `\$12\catcode
  `\&12\catcode `\#12\catcode `\^12\catcode `\_12\catcode `\%12\relax}%
\providecommand \@@startlink[1]{}%
\providecommand \@@endlink[0]{}%
\providecommand \url  [0]{\begingroup\@sanitize@url \@url }%
\providecommand \@url [1]{\endgroup\@href {#1}{\urlprefix }}%
\providecommand \urlprefix  [0]{URL }%
\providecommand \Eprint [0]{\href }%
\providecommand \doibase [0]{http://dx.doi.org/}%
\providecommand \selectlanguage [0]{\@gobble}%
\providecommand \bibinfo  [0]{\@secondoftwo}%
\providecommand \bibfield  [0]{\@secondoftwo}%
\providecommand \translation [1]{[#1]}%
\providecommand \BibitemOpen [0]{}%
\providecommand \bibitemStop [0]{}%
\providecommand \bibitemNoStop [0]{.\EOS\space}%
\providecommand \EOS [0]{\spacefactor3000\relax}%
\providecommand \BibitemShut  [1]{\csname bibitem#1\endcsname}%
\let\auto@bib@innerbib\@empty
\bibitem [{\citenamefont {Skilling}(2004)}]{bib:skilling}%
  \BibitemOpen
  \bibfield  {author} {\bibinfo {author} {\bibfnamefont {J.}~\bibnamefont
  {Skilling}},\ }in\ \href@noop {} {\emph {\bibinfo {booktitle} {AIP Conference
  Proceedings}}},\ Vol.\ \bibinfo {volume} {735}\ (\bibinfo {year} {2004})\ p.\
  \bibinfo {pages} {395}\BibitemShut {NoStop}%
\bibitem [{\citenamefont {Skilling}(2006)}]{bib:skilling2}%
  \BibitemOpen
  \bibfield  {author} {\bibinfo {author} {\bibfnamefont {J.}~\bibnamefont
  {Skilling}},\ }\href@noop {} {\bibfield  {journal} {\bibinfo  {journal} {J.
  of Bayesian Analysis}\ }\textbf {\bibinfo {volume} {1}},\ \bibinfo {pages}
  {833} (\bibinfo {year} {2006})}\BibitemShut {NoStop}%
\bibitem [{\citenamefont {Feroz}\ and\ \citenamefont
  {Hobson}(2008)}]{multinest1}%
  \BibitemOpen
  \bibfield  {author} {\bibinfo {author} {\bibfnamefont {F.}~\bibnamefont
  {Feroz}}\ and\ \bibinfo {author} {\bibfnamefont {M.~P.}\ \bibnamefont
  {Hobson}},\ }\href@noop {} {\bibfield  {journal} {\bibinfo  {journal} {Mon.
  Not. R. Astron. Soc.}\ }\textbf {\bibinfo {volume} {384}},\ \bibinfo {pages}
  {449} (\bibinfo {year} {2008})}\BibitemShut {NoStop}%
\bibitem [{\citenamefont {P\'artay}\ \emph {et~al.}(2010)\citenamefont
  {P\'artay}, \citenamefont {Bart\'ok},\ and\ \citenamefont
  {Cs\'anyi}}]{bib:our_NS_paper}%
  \BibitemOpen
  \bibfield  {author} {\bibinfo {author} {\bibfnamefont {L.~B.}\ \bibnamefont
  {P\'artay}}, \bibinfo {author} {\bibfnamefont {A.~P.}\ \bibnamefont
  {Bart\'ok}}, \ and\ \bibinfo {author} {\bibfnamefont {G.}~\bibnamefont
  {Cs\'anyi}},\ }\href@noop {} {\bibfield  {journal} {\bibinfo  {journal} {J.
  Phys. Chem. B}\ }\textbf {\bibinfo {volume} {114}},\ \bibinfo {pages} {10502}
  (\bibinfo {year} {2010})}\BibitemShut {NoStop}%
\bibitem [{\citenamefont {Burkoff}\ \emph {et~al.}(2012)\citenamefont
  {Burkoff}, \citenamefont {V\'arnai}, \citenamefont {Wells},\ and\
  \citenamefont {Wild}}]{Burkoff1}%
  \BibitemOpen
  \bibfield  {author} {\bibinfo {author} {\bibfnamefont {N.~S.}\ \bibnamefont
  {Burkoff}}, \bibinfo {author} {\bibfnamefont {C.}~\bibnamefont {V\'arnai}},
  \bibinfo {author} {\bibfnamefont {S.~A.}\ \bibnamefont {Wells}}, \ and\
  \bibinfo {author} {\bibfnamefont {D.~L.}\ \bibnamefont {Wild}},\ }\href@noop
  {} {\bibfield  {journal} {\bibinfo  {journal} {Biophys. J.}\ }\textbf
  {\bibinfo {volume} {102}},\ \bibinfo {pages} {878} (\bibinfo {year}
  {2012})}\BibitemShut {NoStop}%
\bibitem [{\citenamefont {Do}\ \emph {et~al.}(2011)\citenamefont {Do},
  \citenamefont {Hirst},\ and\ \citenamefont {Wheatley}}]{bib:wheatley1}%
  \BibitemOpen
  \bibfield  {author} {\bibinfo {author} {\bibfnamefont {H.}~\bibnamefont
  {Do}}, \bibinfo {author} {\bibfnamefont {J.~D.}\ \bibnamefont {Hirst}}, \
  and\ \bibinfo {author} {\bibfnamefont {R.~J.}\ \bibnamefont {Wheatley}},\
  }\href@noop {} {\bibfield  {journal} {\bibinfo  {journal} {J. Chem. Phys.}\
  }\textbf {\bibinfo {volume} {135}},\ \bibinfo {pages} {174105} (\bibinfo
  {year} {2011})}\BibitemShut {NoStop}%
\bibitem [{\citenamefont {Do}\ \emph {et~al.}(2012)\citenamefont {Do},
  \citenamefont {Hirst},\ and\ \citenamefont {Wheatley}}]{bib:wheatley2}%
  \BibitemOpen
  \bibfield  {author} {\bibinfo {author} {\bibfnamefont {H.}~\bibnamefont
  {Do}}, \bibinfo {author} {\bibfnamefont {J.~D.}\ \bibnamefont {Hirst}}, \
  and\ \bibinfo {author} {\bibfnamefont {R.~J.}\ \bibnamefont {Wheatley}},\
  }\href@noop {} {\bibfield  {journal} {\bibinfo  {journal} {J. Phys. Chem.}\
  }\textbf {\bibinfo {volume} {116}},\ \bibinfo {pages} {4535} (\bibinfo {year}
  {2012})}\BibitemShut {NoStop}%
\bibitem [{\citenamefont {Do}\ \emph {et~al.}(2013)\citenamefont {Do}, ,\ and\
  \citenamefont {Wheatley}}]{bib:wheatley3}%
  \BibitemOpen
  \bibfield  {author} {\bibinfo {author} {\bibfnamefont {H.}~\bibnamefont
  {Do}}, , \ and\ \bibinfo {author} {\bibfnamefont {R.~J.}\ \bibnamefont
  {Wheatley}},\ }\href@noop {} {\bibfield  {journal} {\bibinfo  {journal} {J.
  Chem. Theory Comput.}\ }\textbf {\bibinfo {volume} {9}},\ \bibinfo {pages}
  {165} (\bibinfo {year} {2013})}\BibitemShut {NoStop}%
\bibitem [{\citenamefont {Nielsen}(2013)}]{bib:NS_NVT}%
  \BibitemOpen
  \bibfield  {author} {\bibinfo {author} {\bibfnamefont {S.~O.}\ \bibnamefont
  {Nielsen}},\ }\href@noop {} {\bibfield  {journal} {\bibinfo  {journal} {J.
  Chem. Phys.}\ }\textbf {\bibinfo {volume} {139}},\ \bibinfo {pages} {124104}
  (\bibinfo {year} {2013})}\BibitemShut {NoStop}%
\bibitem [{\citenamefont {Skilling}(2009)}]{bib:skilling_convergence}%
  \BibitemOpen
  \bibfield  {author} {\bibinfo {author} {\bibfnamefont {J.}~\bibnamefont
  {Skilling}},\ }in\ \href@noop {} {\emph {\bibinfo {booktitle} {AIP Conference
  Proceedings}}},\ Vol.\ \bibinfo {volume} {1193}\ (\bibinfo {year} {2009})\
  p.\ \bibinfo {pages} {277}\BibitemShut {NoStop}%
\bibitem [{\citenamefont {Rosenbluth}\ and\ \citenamefont
  {Rosenbluth}(1954)}]{bib:HS_Rosenbluth}%
  \BibitemOpen
  \bibfield  {author} {\bibinfo {author} {\bibfnamefont {M.~N.}\ \bibnamefont
  {Rosenbluth}}\ and\ \bibinfo {author} {\bibfnamefont {A.~W.}\ \bibnamefont
  {Rosenbluth}},\ }\href@noop {} {\bibfield  {journal} {\bibinfo  {journal} {J.
  Chem. Phys.}\ }\textbf {\bibinfo {volume} {22}},\ \bibinfo {pages} {881}
  (\bibinfo {year} {1954})}\BibitemShut {NoStop}%
\bibitem [{\citenamefont {Wood}\ and\ \citenamefont
  {Jacobson}(1957)}]{bib:HS_Wood}%
  \BibitemOpen
  \bibfield  {author} {\bibinfo {author} {\bibfnamefont {W.~W.}\ \bibnamefont
  {Wood}}\ and\ \bibinfo {author} {\bibfnamefont {J.~D.}\ \bibnamefont
  {Jacobson}},\ }\href@noop {} {\bibfield  {journal} {\bibinfo  {journal} {J.
  Chem. Phys.}\ }\textbf {\bibinfo {volume} {27}},\ \bibinfo {pages} {1207}
  (\bibinfo {year} {1957})}\BibitemShut {NoStop}%
\bibitem [{\citenamefont {Alder}\ and\ \citenamefont
  {Wainwright}(1957)}]{bib:HS_Wainwright}%
  \BibitemOpen
  \bibfield  {author} {\bibinfo {author} {\bibfnamefont {B.~J.}\ \bibnamefont
  {Alder}}\ and\ \bibinfo {author} {\bibfnamefont {T.~E.}\ \bibnamefont
  {Wainwright}},\ }\href@noop {} {\bibfield  {journal} {\bibinfo  {journal} {J.
  Chem. Phys.}\ }\textbf {\bibinfo {volume} {27}},\ \bibinfo {pages} {1208}
  (\bibinfo {year} {1957})}\BibitemShut {NoStop}%
\bibitem [{\citenamefont {Hoover}\ and\ \citenamefont
  {Ree}(1968)}]{bib:HS_phase_transition}%
  \BibitemOpen
  \bibfield  {author} {\bibinfo {author} {\bibfnamefont {W.~G.}\ \bibnamefont
  {Hoover}}\ and\ \bibinfo {author} {\bibfnamefont {F.~H.}\ \bibnamefont
  {Ree}},\ }\href@noop {} {\bibfield  {journal} {\bibinfo  {journal} {J. Chem.
  Phys.}\ }\textbf {\bibinfo {volume} {49}},\ \bibinfo {pages} {3609} (\bibinfo
  {year} {1968})}\BibitemShut {NoStop}%
\bibitem [{\citenamefont {Wang}\ and\ \citenamefont {Cheung}(2012)}]{HS_bio}%
  \BibitemOpen
  \bibfield  {author} {\bibinfo {author} {\bibfnamefont {Q.}~\bibnamefont
  {Wang}}\ and\ \bibinfo {author} {\bibfnamefont {M.~S.}\ \bibnamefont
  {Cheung}},\ }\href@noop {} {\bibfield  {journal} {\bibinfo  {journal}
  {Biophys. J.}\ }\textbf {\bibinfo {volume} {102}},\ \bibinfo {pages} {2353}
  (\bibinfo {year} {2012})}\BibitemShut {NoStop}%
\bibitem [{\citenamefont {Ishikawa}\ and\ \citenamefont
  {Pedley}(2007)}]{HS_bio2}%
  \BibitemOpen
  \bibfield  {author} {\bibinfo {author} {\bibfnamefont {T.}~\bibnamefont
  {Ishikawa}}\ and\ \bibinfo {author} {\bibfnamefont {T.~J.}\ \bibnamefont
  {Pedley}},\ }\href@noop {} {\bibfield  {journal} {\bibinfo  {journal} {J.
  Fluid Mech.}\ }\textbf {\bibinfo {volume} {588}},\ \bibinfo {pages} {399}
  (\bibinfo {year} {2007})}\BibitemShut {NoStop}%
\bibitem [{\citenamefont {{A. Najafi and R. Golestanian}}(2004)}]{HS_eng}%
  \BibitemOpen
  \bibfield  {author} {\bibinfo {author} {\bibnamefont {{A. Najafi and R.
  Golestanian}}},\ }\href@noop {} {\bibfield  {journal} {\bibinfo  {journal}
  {{Phys. Rev. E}}\ }\textbf {\bibinfo {volume} {{69}}},\ \bibinfo {pages}
  {{062901}} (\bibinfo {year} {{2004}})}\BibitemShut {NoStop}%
\bibitem [{\citenamefont {Torquato}\ and\ \citenamefont
  {Lado}(1986)}]{Composit_model}%
  \BibitemOpen
  \bibfield  {author} {\bibinfo {author} {\bibfnamefont {S.}~\bibnamefont
  {Torquato}}\ and\ \bibinfo {author} {\bibfnamefont {F.}~\bibnamefont
  {Lado}},\ }\href@noop {} {\bibfield  {journal} {\bibinfo  {journal} {Phys.
  Rev. B}\ }\textbf {\bibinfo {volume} {33}},\ \bibinfo {pages} {6428}
  (\bibinfo {year} {1986})}\BibitemShut {NoStop}%
\bibitem [{\citenamefont {Smith}\ and\ \citenamefont
  {Henderson}(1970)}]{HS_anal}%
  \BibitemOpen
  \bibfield  {author} {\bibinfo {author} {\bibfnamefont {W.~R.}\ \bibnamefont
  {Smith}}\ and\ \bibinfo {author} {\bibfnamefont {D.}~\bibnamefont
  {Henderson}},\ }\href@noop {} {\bibfield  {journal} {\bibinfo  {journal}
  {Mol. Phys.}\ }\textbf {\bibinfo {volume} {19}},\ \bibinfo {pages} {411}
  (\bibinfo {year} {1970})}\BibitemShut {NoStop}%
\bibitem [{\citenamefont {Davidchack}\ and\ \citenamefont
  {Laird}(1998)}]{bib:HS_interface}%
  \BibitemOpen
  \bibfield  {author} {\bibinfo {author} {\bibfnamefont {R.~L.}\ \bibnamefont
  {Davidchack}}\ and\ \bibinfo {author} {\bibfnamefont {B.~B.}\ \bibnamefont
  {Laird}},\ }\href@noop {} {\bibfield  {journal} {\bibinfo  {journal} {J.
  Chem. Phys.}\ }\textbf {\bibinfo {volume} {108}},\ \bibinfo {pages} {9452}
  (\bibinfo {year} {1998})}\BibitemShut {NoStop}%
\bibitem [{\citenamefont {Pronk}\ and\ \citenamefont
  {Frenkel}(1999)}]{bib:HS_stacking_faults}%
  \BibitemOpen
  \bibfield  {author} {\bibinfo {author} {\bibfnamefont {S.}~\bibnamefont
  {Pronk}}\ and\ \bibinfo {author} {\bibfnamefont {D.}~\bibnamefont
  {Frenkel}},\ }\href@noop {} {\bibfield  {journal} {\bibinfo  {journal} {J.
  Chem. Phys.}\ }\textbf {\bibinfo {volume} {110}},\ \bibinfo {pages} {4589}
  (\bibinfo {year} {1999})}\BibitemShut {NoStop}%
\bibitem [{\citenamefont {Noya}\ \emph {et~al.}(2008)\citenamefont {Noya},
  \citenamefont {Vega},\ and\ \citenamefont
  {de~Miguel}}]{bib:HS_coexistence_Noya}%
  \BibitemOpen
  \bibfield  {author} {\bibinfo {author} {\bibfnamefont {E.~G.}\ \bibnamefont
  {Noya}}, \bibinfo {author} {\bibfnamefont {C.}~\bibnamefont {Vega}}, \ and\
  \bibinfo {author} {\bibfnamefont {E.}~\bibnamefont {de~Miguel}},\ }\href@noop
  {} {\bibfield  {journal} {\bibinfo  {journal} {J. Chem. Phys.}\ }\textbf
  {\bibinfo {volume} {128}},\ \bibinfo {pages} {154507} (\bibinfo {year}
  {2008})}\BibitemShut {NoStop}%
\bibitem [{\citenamefont {Sanz}\ \emph {et~al.}(2011)\citenamefont {Sanz},
  \citenamefont {Valeriani}, \citenamefont {Zaccarelli}, \citenamefont {Poon},
  \citenamefont {Pusey},\ and\ \citenamefont {Cates}}]{bib:HS_crystallization}%
  \BibitemOpen
  \bibfield  {author} {\bibinfo {author} {\bibfnamefont {E.}~\bibnamefont
  {Sanz}}, \bibinfo {author} {\bibfnamefont {C.}~\bibnamefont {Valeriani}},
  \bibinfo {author} {\bibfnamefont {E.}~\bibnamefont {Zaccarelli}}, \bibinfo
  {author} {\bibfnamefont {W.~C.~K.}\ \bibnamefont {Poon}}, \bibinfo {author}
  {\bibfnamefont {P.~N.}\ \bibnamefont {Pusey}}, \ and\ \bibinfo {author}
  {\bibfnamefont {M.~E.}\ \bibnamefont {Cates}},\ }\href@noop {} {\bibfield
  {journal} {\bibinfo  {journal} {Phys. Rev. Lett.}\ }\textbf {\bibinfo
  {volume} {106}},\ \bibinfo {pages} {215701} (\bibinfo {year}
  {2011})}\BibitemShut {NoStop}%
\bibitem [{\citenamefont {Schilling}\ \emph {et~al.}(2010)\citenamefont
  {Schilling}, \citenamefont {Sch\"ope}, \citenamefont {Oettel}, \citenamefont
  {Opletal},\ and\ \citenamefont {Snook}}]{bib:HS_crystallization2}%
  \BibitemOpen
  \bibfield  {author} {\bibinfo {author} {\bibfnamefont {T.}~\bibnamefont
  {Schilling}}, \bibinfo {author} {\bibfnamefont {H.~J.}\ \bibnamefont
  {Sch\"ope}}, \bibinfo {author} {\bibfnamefont {M.}~\bibnamefont {Oettel}},
  \bibinfo {author} {\bibfnamefont {G.}~\bibnamefont {Opletal}}, \ and\
  \bibinfo {author} {\bibfnamefont {I.}~\bibnamefont {Snook}},\ }\href@noop {}
  {\bibfield  {journal} {\bibinfo  {journal} {Phys. Rev. Lett.}\ }\textbf
  {\bibinfo {volume} {105}},\ \bibinfo {pages} {025701} (\bibinfo {year}
  {2010})}\BibitemShut {NoStop}%
\bibitem [{\citenamefont {Williams}\ \emph {et~al.}(2001)\citenamefont
  {Williams}, \citenamefont {Snook},\ and\ \citenamefont {van
  Megen}}]{bib:HS_noglass}%
  \BibitemOpen
  \bibfield  {author} {\bibinfo {author} {\bibfnamefont {S.~R.}\ \bibnamefont
  {Williams}}, \bibinfo {author} {\bibfnamefont {I.~K.}\ \bibnamefont {Snook}},
  \ and\ \bibinfo {author} {\bibfnamefont {W.}~\bibnamefont {van Megen}},\
  }\href@noop {} {\bibfield  {journal} {\bibinfo  {journal} {Phys. Rev. E}\
  }\textbf {\bibinfo {volume} {64}},\ \bibinfo {pages} {021506} (\bibinfo
  {year} {2001})}\BibitemShut {NoStop}%
\bibitem [{\citenamefont {Alder}\ \emph {et~al.}(1968)\citenamefont {Alder},
  \citenamefont {Hoover},\ and\ \citenamefont {Young}}]{EoS_Alder}%
  \BibitemOpen
  \bibfield  {author} {\bibinfo {author} {\bibfnamefont {B.~J.}\ \bibnamefont
  {Alder}}, \bibinfo {author} {\bibfnamefont {W.~G.}\ \bibnamefont {Hoover}}, \
  and\ \bibinfo {author} {\bibfnamefont {D.~A.}\ \bibnamefont {Young}},\
  }\href@noop {} {\bibfield  {journal} {\bibinfo  {journal} {The Journal of
  chemical physics}\ }\textbf {\bibinfo {volume} {49}},\ \bibinfo {pages}
  {3688} (\bibinfo {year} {1968})}\BibitemShut {NoStop}%
\bibitem [{\citenamefont {Santos}\ and\ \citenamefont {L{\'o}pez~de
  Haro}(2009)}]{EoS_Santos}%
  \BibitemOpen
  \bibfield  {author} {\bibinfo {author} {\bibfnamefont {A.}~\bibnamefont
  {Santos}}\ and\ \bibinfo {author} {\bibfnamefont {M.}~\bibnamefont
  {L{\'o}pez~de Haro}},\ }\href@noop {} {\bibfield  {journal} {\bibinfo
  {journal} {The Journal of chemical physics}\ }\textbf {\bibinfo {volume}
  {130}},\ \bibinfo {pages} {214104} (\bibinfo {year} {2009})}\BibitemShut
  {NoStop}%
\bibitem [{\citenamefont {Wang}\ \emph {et~al.}(1996)\citenamefont {Wang},
  \citenamefont {Khoshkbarchi},\ and\ \citenamefont {Vera}}]{EoS_Wang}%
  \BibitemOpen
  \bibfield  {author} {\bibinfo {author} {\bibfnamefont {W.}~\bibnamefont
  {Wang}}, \bibinfo {author} {\bibfnamefont {M.~K.}\ \bibnamefont
  {Khoshkbarchi}}, \ and\ \bibinfo {author} {\bibfnamefont {J.~H.}\
  \bibnamefont {Vera}},\ }\href@noop {} {\bibfield  {journal} {\bibinfo
  {journal} {Fluid phase equilibria}\ }\textbf {\bibinfo {volume} {115}},\
  \bibinfo {pages} {25} (\bibinfo {year} {1996})}\BibitemShut {NoStop}%
\bibitem [{\citenamefont {Bannerman}\ \emph {et~al.}(2010)\citenamefont
  {Bannerman}, \citenamefont {Lue},\ and\ \citenamefont {Woodcock}}]{EoS_WC1}%
  \BibitemOpen
  \bibfield  {author} {\bibinfo {author} {\bibfnamefont {M.~N.}\ \bibnamefont
  {Bannerman}}, \bibinfo {author} {\bibfnamefont {L.}~\bibnamefont {Lue}}, \
  and\ \bibinfo {author} {\bibfnamefont {L.~V.}\ \bibnamefont {Woodcock}},\
  }\href@noop {} {\bibfield  {journal} {\bibinfo  {journal} {The Journal of
  chemical physics}\ }\textbf {\bibinfo {volume} {132}},\ \bibinfo {pages}
  {084507} (\bibinfo {year} {2010})}\BibitemShut {NoStop}%
\bibitem [{\citenamefont {Bolhuis}\ \emph {et~al.}(1997)\citenamefont
  {Bolhuis}, \citenamefont {Frenkel}, \citenamefont {Mau},\ and\ \citenamefont
  {Huse}}]{Bolhuis_fcc-hcp}%
  \BibitemOpen
  \bibfield  {author} {\bibinfo {author} {\bibfnamefont {P.~G.}\ \bibnamefont
  {Bolhuis}}, \bibinfo {author} {\bibfnamefont {D.}~\bibnamefont {Frenkel}},
  \bibinfo {author} {\bibfnamefont {S.-C.}\ \bibnamefont {Mau}}, \ and\
  \bibinfo {author} {\bibfnamefont {D.~A.}\ \bibnamefont {Huse}},\ }\href@noop
  {} {\bibfield  {journal} {\bibinfo  {journal} {Nature}\ }\textbf {\bibinfo
  {volume} {388}},\ \bibinfo {pages} {235} (\bibinfo {year}
  {1997})}\BibitemShut {NoStop}%
\bibitem [{\citenamefont {Pusey}\ and\ \citenamefont {van
  Megen}(1986)}]{HS_exp_PMMA}%
  \BibitemOpen
  \bibfield  {author} {\bibinfo {author} {\bibfnamefont {P.~N.}\ \bibnamefont
  {Pusey}}\ and\ \bibinfo {author} {\bibfnamefont {W.}~\bibnamefont {van
  Megen}},\ }\href@noop {} {\bibfield  {journal} {\bibinfo  {journal} {Nature}\
  }\textbf {\bibinfo {volume} {320}},\ \bibinfo {pages} {340} (\bibinfo {year}
  {1986})}\BibitemShut {NoStop}%
\bibitem [{\citenamefont {Sch\"ope}\ \emph {et~al.}(2006)\citenamefont
  {Sch\"ope}, \citenamefont {Bryant},\ and\ \citenamefont {van
  Megen}}]{bib:HS_crystallization_exp}%
  \BibitemOpen
  \bibfield  {author} {\bibinfo {author} {\bibfnamefont {H.~J.}\ \bibnamefont
  {Sch\"ope}}, \bibinfo {author} {\bibfnamefont {G.}~\bibnamefont {Bryant}}, \
  and\ \bibinfo {author} {\bibfnamefont {W.}~\bibnamefont {van Megen}},\
  }\href@noop {} {\bibfield  {journal} {\bibinfo  {journal} {Phys. Rev. Lett.}\
  }\textbf {\bibinfo {volume} {96}},\ \bibinfo {pages} {175701} (\bibinfo
  {year} {2006})}\BibitemShut {NoStop}%
\bibitem [{\citenamefont {Sch\"atzel}\ and\ \citenamefont
  {Ackerson}(1993)}]{bib:HS_crystallization_exp2}%
  \BibitemOpen
  \bibfield  {author} {\bibinfo {author} {\bibfnamefont {K.}~\bibnamefont
  {Sch\"atzel}}\ and\ \bibinfo {author} {\bibfnamefont {B.~J.}\ \bibnamefont
  {Ackerson}},\ }\href@noop {} {\bibfield  {journal} {\bibinfo  {journal}
  {Phys. Rev. E}\ }\textbf {\bibinfo {volume} {48}},\ \bibinfo {pages} {3766}
  (\bibinfo {year} {1993})}\BibitemShut {NoStop}%
\bibitem [{\citenamefont {Harland}\ and\ \citenamefont {van
  Megen}(1997)}]{bib:HS_crystallization_exp3}%
  \BibitemOpen
  \bibfield  {author} {\bibinfo {author} {\bibfnamefont {J.~L.}\ \bibnamefont
  {Harland}}\ and\ \bibinfo {author} {\bibfnamefont {W.}~\bibnamefont {van
  Megen}},\ }\href@noop {} {\bibfield  {journal} {\bibinfo  {journal} {Phys.
  Rev. E}\ }\textbf {\bibinfo {volume} {55}},\ \bibinfo {pages} {3054}
  (\bibinfo {year} {1997})}\BibitemShut {NoStop}%
\bibitem [{\citenamefont {Cheng}\ \emph
  {et~al.}(2001{\natexlab{a}})\citenamefont {Cheng}, \citenamefont {Chaikin},
  \citenamefont {Russel}, \citenamefont {Meyer}, \citenamefont {Zhu},
  \citenamefont {Rogers},\ and\ \citenamefont
  {Ottewill}}]{HS_exp_spaceshuttle}%
  \BibitemOpen
  \bibfield  {author} {\bibinfo {author} {\bibfnamefont {Z.}~\bibnamefont
  {Cheng}}, \bibinfo {author} {\bibfnamefont {P.}~\bibnamefont {Chaikin}},
  \bibinfo {author} {\bibfnamefont {W.}~\bibnamefont {Russel}}, \bibinfo
  {author} {\bibfnamefont {W.}~\bibnamefont {Meyer}}, \bibinfo {author}
  {\bibfnamefont {J.}~\bibnamefont {Zhu}}, \bibinfo {author} {\bibfnamefont
  {R.}~\bibnamefont {Rogers}}, \ and\ \bibinfo {author} {\bibfnamefont
  {R.}~\bibnamefont {Ottewill}},\ }\href@noop {} {\bibfield  {journal}
  {\bibinfo  {journal} {Materials \& Design}\ }\textbf {\bibinfo {volume}
  {22}},\ \bibinfo {pages} {529} (\bibinfo {year}
  {2001}{\natexlab{a}})}\BibitemShut {NoStop}%
\bibitem [{\citenamefont {Zhu}\ \emph {et~al.}(1997)\citenamefont {Zhu},
  \citenamefont {Li}, \citenamefont {Rogers}, \citenamefont {Meyer},
  \citenamefont {Ottewill}, \citenamefont {Russel},\ and\ \citenamefont
  {Chaikin}}]{Zhu_spaceshuttle}%
  \BibitemOpen
  \bibfield  {author} {\bibinfo {author} {\bibfnamefont {J.}~\bibnamefont
  {Zhu}}, \bibinfo {author} {\bibfnamefont {M.}~\bibnamefont {Li}}, \bibinfo
  {author} {\bibfnamefont {R.}~\bibnamefont {Rogers}}, \bibinfo {author}
  {\bibfnamefont {W.}~\bibnamefont {Meyer}}, \bibinfo {author} {\bibfnamefont
  {R.~H.}\ \bibnamefont {Ottewill}}, \bibinfo {author} {\bibfnamefont {W.~B.}\
  \bibnamefont {Russel}}, \ and\ \bibinfo {author} {\bibfnamefont {P.~M.}\
  \bibnamefont {Chaikin}},\ }\href@noop {} {\bibfield  {journal} {\bibinfo
  {journal} {Nature}\ }\textbf {\bibinfo {volume} {387}},\ \bibinfo {pages}
  {883} (\bibinfo {year} {1997})}\BibitemShut {NoStop}%
\bibitem [{\citenamefont {Petukhov}\ \emph {et~al.}(2003)\citenamefont
  {Petukhov}, \citenamefont {Dolbnya}, \citenamefont {Aarts}, \citenamefont
  {Vroege},\ and\ \citenamefont {Lekkerkerker}}]{Lekker_diff}%
  \BibitemOpen
  \bibfield  {author} {\bibinfo {author} {\bibfnamefont {A.}~\bibnamefont
  {Petukhov}}, \bibinfo {author} {\bibfnamefont {I.}~\bibnamefont {Dolbnya}},
  \bibinfo {author} {\bibfnamefont {D.}~\bibnamefont {Aarts}}, \bibinfo
  {author} {\bibfnamefont {G.}~\bibnamefont {Vroege}}, \ and\ \bibinfo {author}
  {\bibfnamefont {H.}~\bibnamefont {Lekkerkerker}},\ }\href@noop {} {\bibfield
  {journal} {\bibinfo  {journal} {Physical Review Letters}\ }\textbf {\bibinfo
  {volume} {90}},\ \bibinfo {pages} {028304} (\bibinfo {year}
  {2003})}\BibitemShut {NoStop}%
\bibitem [{\citenamefont {Hales}(2005)}]{bib:Keplers_proof}%
  \BibitemOpen
  \bibfield  {author} {\bibinfo {author} {\bibfnamefont {T.~C.}\ \bibnamefont
  {Hales}},\ }\href@noop {} {\bibfield  {journal} {\bibinfo  {journal} {Ann.
  Math.}\ }\textbf {\bibinfo {volume} {162}},\ \bibinfo {pages} {1065}
  (\bibinfo {year} {2005})}\BibitemShut {NoStop}%
\bibitem [{\citenamefont {Rintoul}\ and\ \citenamefont
  {Torquato}(1996)}]{bib:HS_noglass_gr}%
  \BibitemOpen
  \bibfield  {author} {\bibinfo {author} {\bibfnamefont {M.~D.}\ \bibnamefont
  {Rintoul}}\ and\ \bibinfo {author} {\bibfnamefont {S.}~\bibnamefont
  {Torquato}},\ }\href@noop {} {\bibfield  {journal} {\bibinfo  {journal}
  {Phys. Rev. Lett.}\ }\textbf {\bibinfo {volume} {77}},\ \bibinfo {pages}
  {4198} (\bibinfo {year} {1996})}\BibitemShut {NoStop}%
\bibitem [{\citenamefont {Torquato}\ \emph {et~al.}(2000)\citenamefont
  {Torquato}, \citenamefont {Truskett},\ and\ \citenamefont
  {Debenedetti}}]{bib:debenedetti_RCP}%
  \BibitemOpen
  \bibfield  {author} {\bibinfo {author} {\bibfnamefont {S.}~\bibnamefont
  {Torquato}}, \bibinfo {author} {\bibfnamefont {T.~M.}\ \bibnamefont
  {Truskett}}, \ and\ \bibinfo {author} {\bibfnamefont {P.~G.}\ \bibnamefont
  {Debenedetti}},\ }\href@noop {} {\bibfield  {journal} {\bibinfo  {journal}
  {Phys. Rev. Lett.}\ }\textbf {\bibinfo {volume} {84}},\ \bibinfo {pages}
  {2064} (\bibinfo {year} {2000})}\BibitemShut {NoStop}%
\bibitem [{\citenamefont {Scott}\ and\ \citenamefont
  {Kilgour}(1969)}]{bib:RCP_exp}%
  \BibitemOpen
  \bibfield  {author} {\bibinfo {author} {\bibfnamefont {G.}~\bibnamefont
  {Scott}}\ and\ \bibinfo {author} {\bibfnamefont {D.}~\bibnamefont
  {Kilgour}},\ }\href@noop {} {\bibfield  {journal} {\bibinfo  {journal} {Br.
  J. Appl. Phys.}\ }\textbf {\bibinfo {volume} {2}},\ \bibinfo {pages} {863}
  (\bibinfo {year} {1969})}\BibitemShut {NoStop}%
\bibitem [{\citenamefont {Pouliquen}\ \emph {et~al.}(1997)\citenamefont
  {Pouliquen}, \citenamefont {Nicolas},\ and\ \citenamefont
  {Weidman}}]{bib:RCP_exp2}%
  \BibitemOpen
  \bibfield  {author} {\bibinfo {author} {\bibfnamefont {O.}~\bibnamefont
  {Pouliquen}}, \bibinfo {author} {\bibfnamefont {M.}~\bibnamefont {Nicolas}},
  \ and\ \bibinfo {author} {\bibfnamefont {P.~D.}\ \bibnamefont {Weidman}},\
  }\href@noop {} {\bibfield  {journal} {\bibinfo  {journal} {Phys. Rev. Lett.}\
  }\textbf {\bibinfo {volume} {79}},\ \bibinfo {pages} {3640} (\bibinfo {year}
  {1997})}\BibitemShut {NoStop}%
\bibitem [{\citenamefont {Jodrey}\ and\ \citenamefont
  {Tory}(1985)}]{bib:RCP_calc_densification}%
  \BibitemOpen
  \bibfield  {author} {\bibinfo {author} {\bibfnamefont {W.~S.}\ \bibnamefont
  {Jodrey}}\ and\ \bibinfo {author} {\bibfnamefont {E.~M.}\ \bibnamefont
  {Tory}},\ }\href@noop {} {\bibfield  {journal} {\bibinfo  {journal} {Phys.
  Rev. A}\ }\textbf {\bibinfo {volume} {32}},\ \bibinfo {pages} {2347}
  (\bibinfo {year} {1985})}\BibitemShut {NoStop}%
\bibitem [{\citenamefont {Tobochnik}\ and\ \citenamefont
  {Chapin}(1988)}]{bib:RCP_calc_MC}%
  \BibitemOpen
  \bibfield  {author} {\bibinfo {author} {\bibfnamefont {J.}~\bibnamefont
  {Tobochnik}}\ and\ \bibinfo {author} {\bibfnamefont {P.}~\bibnamefont
  {Chapin}},\ }\href@noop {} {\bibfield  {journal} {\bibinfo  {journal} {J.
  Chem. Phys.}\ }\textbf {\bibinfo {volume} {88}},\ \bibinfo {pages} {5824}
  (\bibinfo {year} {1988})}\BibitemShut {NoStop}%
\bibitem [{\citenamefont {Visscher}\ and\ \citenamefont
  {Bolsterli}(1972)}]{bib:RCP_calc_dropandroll}%
  \BibitemOpen
  \bibfield  {author} {\bibinfo {author} {\bibfnamefont {W.~M.}\ \bibnamefont
  {Visscher}}\ and\ \bibinfo {author} {\bibfnamefont {M.}~\bibnamefont
  {Bolsterli}},\ }\href@noop {} {\bibfield  {journal} {\bibinfo  {journal}
  {Nature}\ }\textbf {\bibinfo {volume} {239}},\ \bibinfo {pages} {504}
  (\bibinfo {year} {1972})}\BibitemShut {NoStop}%
\bibitem [{\citenamefont {Torquato}\ and\ \citenamefont
  {Stillinger}(2001)}]{bib:torquato_jamming}%
  \BibitemOpen
  \bibfield  {author} {\bibinfo {author} {\bibfnamefont {S.}~\bibnamefont
  {Torquato}}\ and\ \bibinfo {author} {\bibfnamefont {F.~H.}\ \bibnamefont
  {Stillinger}},\ }\href@noop {} {\bibfield  {journal} {\bibinfo  {journal} {J.
  Phys. Chem. B}\ }\textbf {\bibinfo {volume} {105}},\ \bibinfo {pages} {11849}
  (\bibinfo {year} {2001})}\BibitemShut {NoStop}%
\bibitem [{\citenamefont {Torquato}\ and\ \citenamefont
  {Stillinger}(2010)}]{bib:torquato_RCP}%
  \BibitemOpen
  \bibfield  {author} {\bibinfo {author} {\bibfnamefont {S.}~\bibnamefont
  {Torquato}}\ and\ \bibinfo {author} {\bibfnamefont {F.~H.}\ \bibnamefont
  {Stillinger}},\ }\href@noop {} {\bibfield  {journal} {\bibinfo  {journal}
  {Rev. Mod. Phys.}\ }\textbf {\bibinfo {volume} {82}},\ \bibinfo {pages}
  {2633} (\bibinfo {year} {2010})}\BibitemShut {NoStop}%
\bibitem [{\citenamefont {Mulero}\ \emph {et~al.}(2008)\citenamefont {Mulero},
  \citenamefont {Galan}, \citenamefont {Parra},\ and\ \citenamefont
  {Cuadros}}]{bib:Mulera_HSEoS_sum}%
  \BibitemOpen
  \bibfield  {author} {\bibinfo {author} {\bibfnamefont {A.}~\bibnamefont
  {Mulero}}, \bibinfo {author} {\bibfnamefont {C.~A.}\ \bibnamefont {Galan}},
  \bibinfo {author} {\bibfnamefont {M.~I.}\ \bibnamefont {Parra}}, \ and\
  \bibinfo {author} {\bibfnamefont {F.}~\bibnamefont {Cuadros}},\ }\href@noop
  {} {\emph {\bibinfo {title} {Theory and Simulation of Hard-Sphere Fluids and
  Related Systems}}},\ \bibinfo {series} {Lecture Notes in Physics}, Vol.\
  \bibinfo {volume} {753}\ (\bibinfo  {publisher} {Springer-Verlag, Berlin},\
  \bibinfo {year} {2008})\BibitemShut {NoStop}%
\bibitem [{\citenamefont {Carnahan}\ and\ \citenamefont
  {Starling}(1969)}]{EoS_CS}%
  \BibitemOpen
  \bibfield  {author} {\bibinfo {author} {\bibfnamefont {N.~F.}\ \bibnamefont
  {Carnahan}}\ and\ \bibinfo {author} {\bibfnamefont {K.~E.}\ \bibnamefont
  {Starling}},\ }\href@noop {} {\bibfield  {journal} {\bibinfo  {journal} {J.
  Chem. Phys.}\ }\textbf {\bibinfo {volume} {51}},\ \bibinfo {pages} {635}
  (\bibinfo {year} {1969})}\BibitemShut {NoStop}%
\bibitem [{\citenamefont {Boublik}(1986)}]{EoS_Kolafa}%
  \BibitemOpen
  \bibfield  {author} {\bibinfo {author} {\bibfnamefont {T.}~\bibnamefont
  {Boublik}},\ }\href@noop {} {\bibfield  {journal} {\bibinfo  {journal} {Mol.
  Phys.}\ }\textbf {\bibinfo {volume} {59}},\ \bibinfo {pages} {371} (\bibinfo
  {year} {1986})}\BibitemShut {NoStop}%
\bibitem [{\citenamefont {Speedy}(1998)}]{EoS_Speedy}%
  \BibitemOpen
  \bibfield  {author} {\bibinfo {author} {\bibfnamefont {R.~J.}\ \bibnamefont
  {Speedy}},\ }\href@noop {} {\bibfield  {journal} {\bibinfo  {journal}
  {Journal Of Physics-Condensed Matter}\ }\textbf {\bibinfo {volume} {10}},\
  \bibinfo {pages} {4387} (\bibinfo {year} {1998})}\BibitemShut {NoStop}%
\bibitem [{\citenamefont {Steinhardt}\ \emph {et~al.}(1983)\citenamefont
  {Steinhardt}, \citenamefont {Nelson},\ and\ \citenamefont
  {Ronchetti}}]{bib:Q6parameter}%
  \BibitemOpen
  \bibfield  {author} {\bibinfo {author} {\bibfnamefont {P.~J.}\ \bibnamefont
  {Steinhardt}}, \bibinfo {author} {\bibfnamefont {D.~R.}\ \bibnamefont
  {Nelson}}, \ and\ \bibinfo {author} {\bibfnamefont {M.}~\bibnamefont
  {Ronchetti}},\ }\href@noop {} {\bibfield  {journal} {\bibinfo  {journal}
  {Phys. Rev. B}\ }\textbf {\bibinfo {volume} {28}},\ \bibinfo {pages} {784}
  (\bibinfo {year} {1983})}\BibitemShut {NoStop}%
\bibitem [{\citenamefont {Pronk}\ and\ \citenamefont
  {Frenkel}(2003)}]{Pronk_elastic_diff}%
  \BibitemOpen
  \bibfield  {author} {\bibinfo {author} {\bibfnamefont {S.}~\bibnamefont
  {Pronk}}\ and\ \bibinfo {author} {\bibfnamefont {D.}~\bibnamefont
  {Frenkel}},\ }\href@noop {} {\bibfield  {journal} {\bibinfo  {journal}
  {Physical Review Letters}\ }\textbf {\bibinfo {volume} {90}},\ \bibinfo
  {pages} {255501} (\bibinfo {year} {2003})}\BibitemShut {NoStop}%
\bibitem [{\citenamefont {Woodcock}(1997)}]{Woodcock_hcpfcc}%
  \BibitemOpen
  \bibfield  {author} {\bibinfo {author} {\bibfnamefont {L.~V.}\ \bibnamefont
  {Woodcock}},\ }\href@noop {} {\bibfield  {journal} {\bibinfo  {journal}
  {Nature}\ }\textbf {\bibinfo {volume} {385}},\ \bibinfo {pages} {141}
  (\bibinfo {year} {1997})}\BibitemShut {NoStop}%
\bibitem [{\citenamefont {Cheng}\ \emph
  {et~al.}(2001{\natexlab{b}})\citenamefont {Cheng}, \citenamefont {Chaikin},
  \citenamefont {Zhu}, \citenamefont {Russel},\ and\ \citenamefont
  {Meyer}}]{PhysRevLett.88.015501}%
  \BibitemOpen
  \bibfield  {author} {\bibinfo {author} {\bibfnamefont {Z.}~\bibnamefont
  {Cheng}}, \bibinfo {author} {\bibfnamefont {P.~M.}\ \bibnamefont {Chaikin}},
  \bibinfo {author} {\bibfnamefont {J.}~\bibnamefont {Zhu}}, \bibinfo {author}
  {\bibfnamefont {W.~B.}\ \bibnamefont {Russel}}, \ and\ \bibinfo {author}
  {\bibfnamefont {W.~V.}\ \bibnamefont {Meyer}},\ }\href@noop {} {\bibfield
  {journal} {\bibinfo  {journal} {Phys. Rev. Lett.}\ }\textbf {\bibinfo
  {volume} {88}},\ \bibinfo {pages} {015501} (\bibinfo {year}
  {2001}{\natexlab{b}})}\BibitemShut {NoStop}%
\end{thebibliography}%

\newpage

\section{Appendix}

\subsection{Computational details}

During the Monte Carlo process, both the particle positions and the shape of the simulation box are changed. In each Monte Carlo step either the movement of a single particle or a lattice shape change is proposed (the probability of the latter being $1/(N+1)$). In case of a shape change, the three lattice vectors are perturbed by small random vectors and, in order to avoid very flattened box shapes for which finite size effects would be very large, we applied the constraint that the smallest height of the box may not become less than 55\% of the height of the cubic box at the same volume.
To keep the acceptance ratio between 10\% and 20\% for both the particle position and lattice shape changes, the step sizes of each are readjusted at every 1000 NS iteration. 

When evaluating sums in eq.~\ref{eq:posint} we work in quadruple precision. Alternatively, if using the usual double precision floating point arithmetic one has to be very careful to avoid under- or overflow.

\subsection{Quenching}

\begin{figure}[bthp]
\begin{center}
\includegraphics[width=9.5cm]{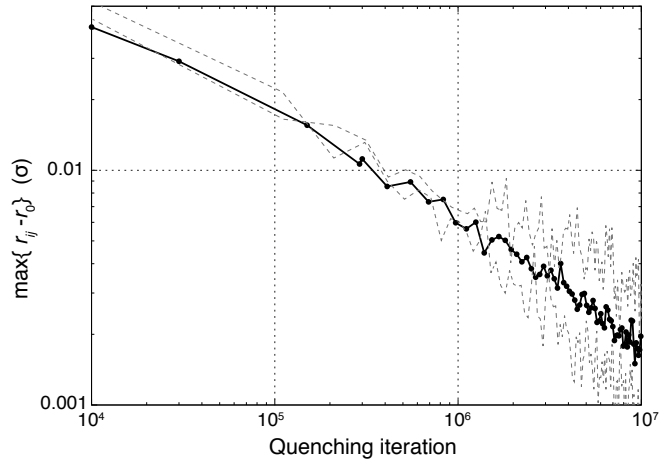}
\end{center}
\caption {Convergence towards jammed states during the quenching process. The difference between the furthest and closest first neighbour distances (the latter is, by definition, the particle diameter) is calculated within every configuration, and plotted against the number of MC steps. The black line is the average of several quenching processes. The grey dashed curves represents the slowest and the fastest process from our sample. Both axes are in the logarithmic scale.}
\label{fig:quench_test}
\end{figure}

Here we give details on the procedure that we used to quench configurations in order to obtain noncrystalline jammed configurations (note that configurations that include ``rattle'' particles are not excluded). During the quenching process the particle size was increased after every MC step until the whole configuration was jammed. We considered a configuration jammed when further relaxation could only increase the packing fraction by less then $0.3\%$ in $10^6$ trial movements of the particles. Figure~\ref{fig:quench_test} shows some evidence that this procedure really does result in configurations that are very close to being jammed: the largest of the nearest-neighbour distances as a function of iteration number approaches the particle diameter approximately according to a power law with an exponent of $-1/2$, as expected for a Monte Carlo process. We have to note that this quenching process produces locally jammed configurations which are not necessarily always collectively jammed too. 

\begin{figure}[tbhp]
\includegraphics[width=5.5cm]{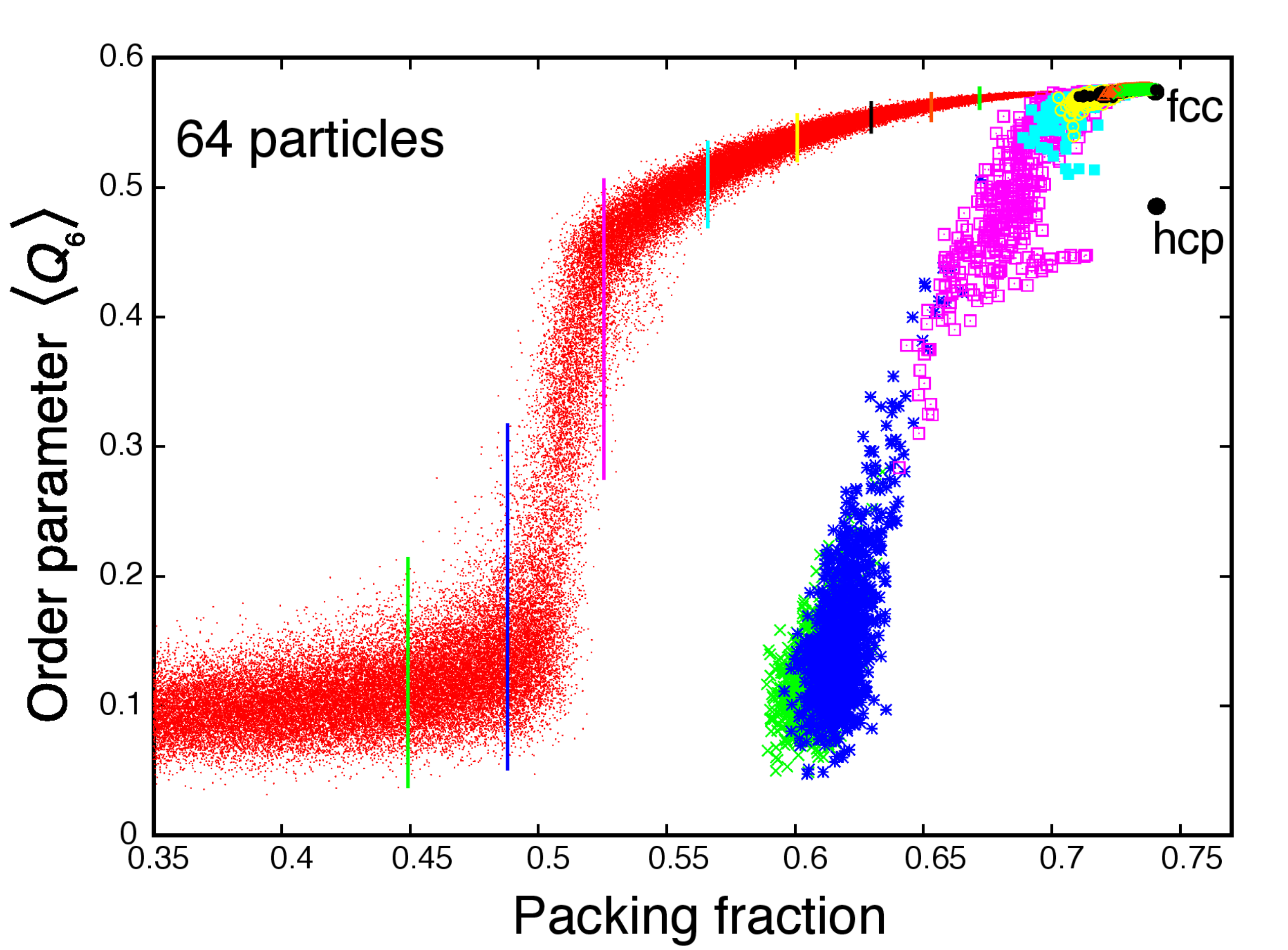}
\includegraphics[width=5.5cm]{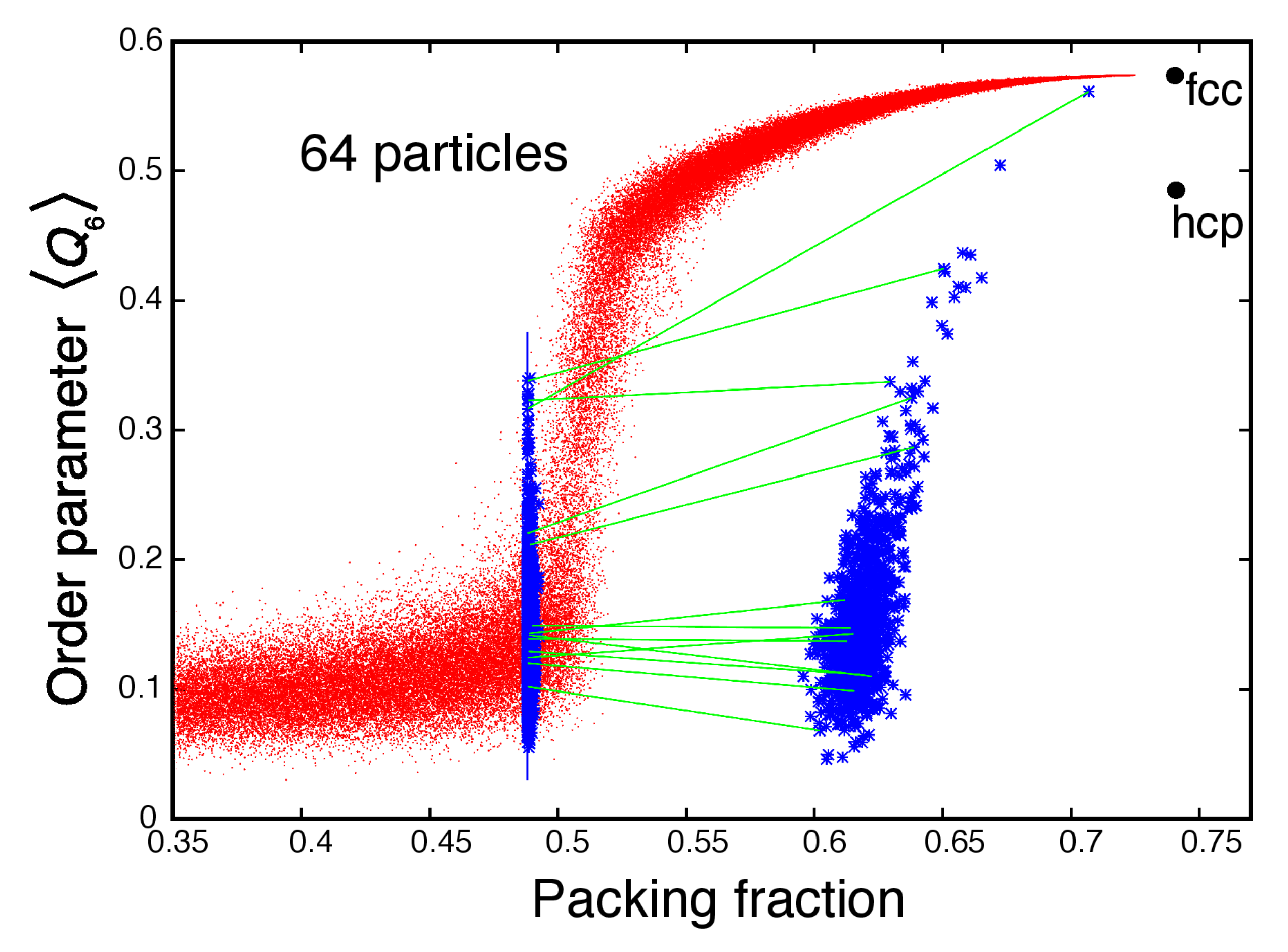}
\includegraphics[width=5.5cm]{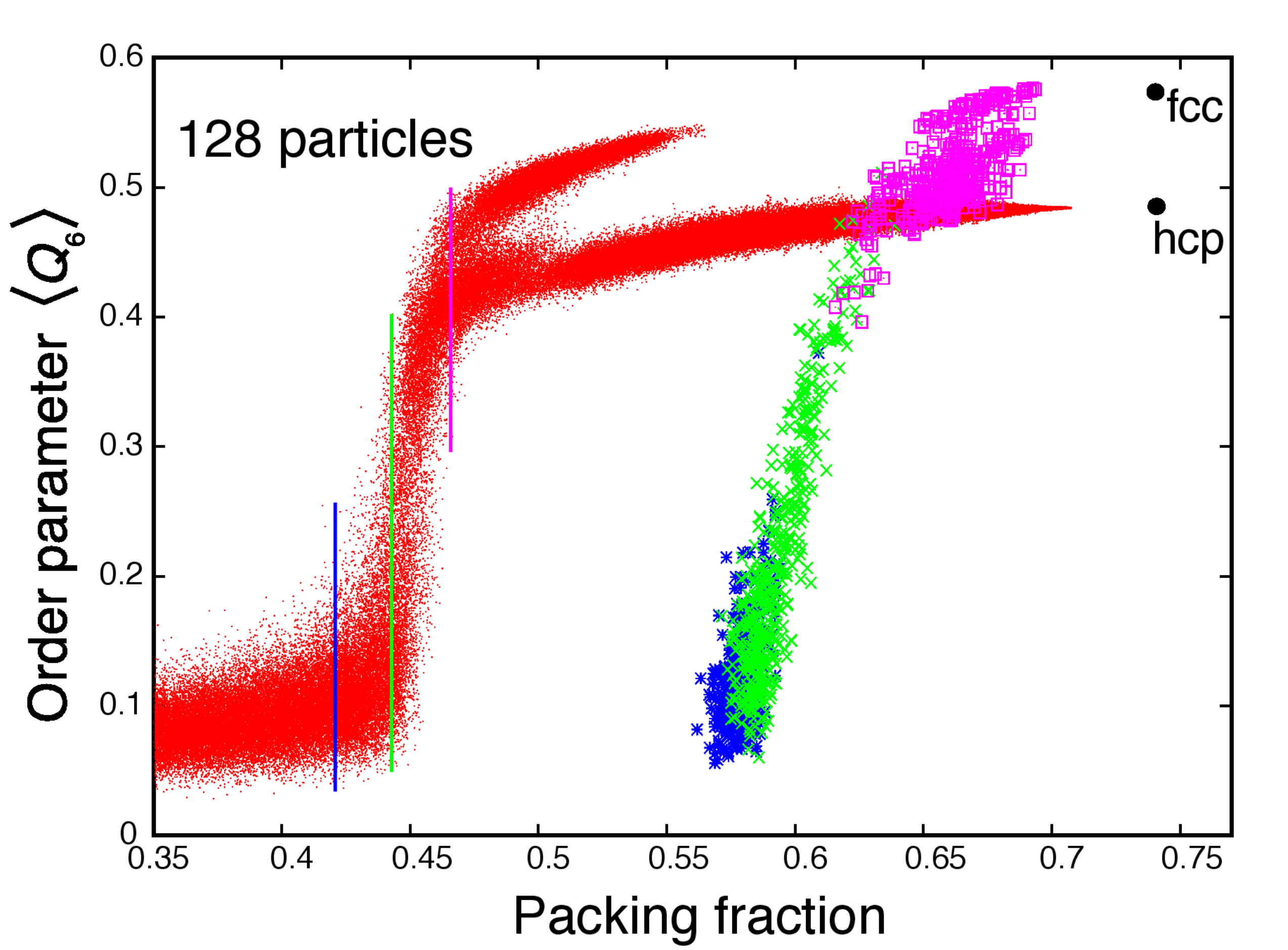}
\includegraphics[width=5.5cm]{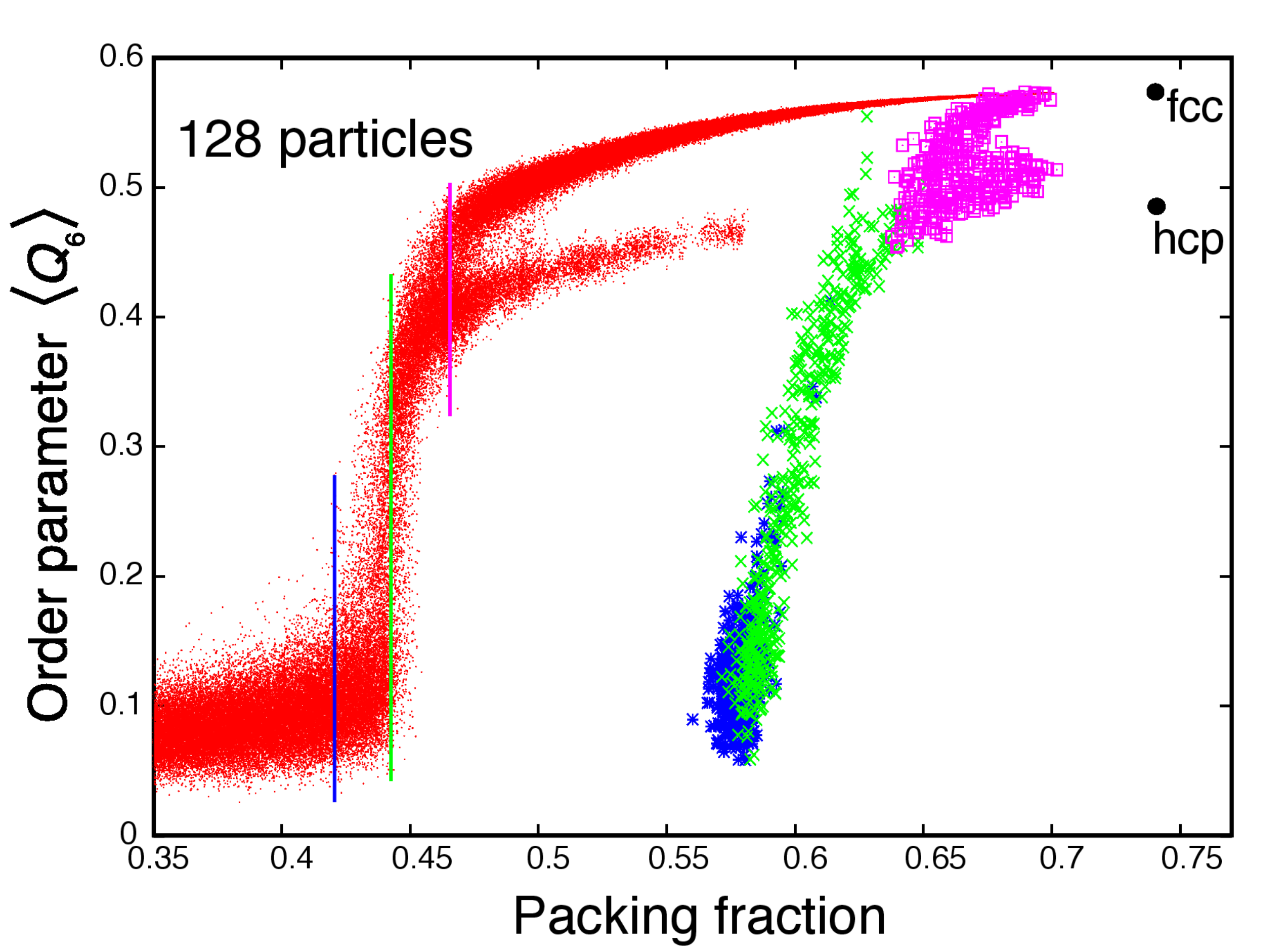}
\caption {The $Q_6$ order parameter as a function of the packing fraction in case of an individual run with 64 hard spheres (top panels) and two runs with 128 hard spheres (bottom panels). The red dots represent the configurations generated during the nested sampling runs. The vertical lines indicate the packing fraction levels at which the sample sets were saved and subsequently quenched in order to obtain the jammed configurations. The quenched jammed configurations are shown by symbols with matching colours. In the top right panel some of the initial configurations and their corresponding jammed structures are connected by green lines to show how the quenching changes the order parameter. The points corresponding to perfect fcc and hcp structures are shown by black circles.}\label{fig:LP_jammed}
\end{figure}

The question arises how the location of a given jammed state on the phase diagram depends on the location of the initial configuration it was quenched from. We chose several packing fraction values and  took the corresponding sample set of the nested sampling run in order to have sets of configurations uniformly distributed in configuration space at given packing fraction levels and we quenched those with the same method as previously described. The chosen packing fraction levels and the quenched jammed states are shown with matching colours in Figure~\ref{fig:LP_jammed} for an individual run with 64 and two runs of the 128 hard sphere systems. The figures show the same general picture for both system sizes. Different initial packing fraction levels result in different groups of jammed states. Starting from the fluid phase ($\phi \approx 0.45$) the jammed states have relatively small packing fraction and low order, while in case of quenching configurations taken from above the freezing point the jammed states end up in the region of crystalline structures. 

This suggests that around the freezing point the relative phase space volume of the crystalline structures is small, approximately only $0.2\%$ of the points quenched from the set with $\phi \approx 0.485$ ended up with a high order parameter. Thus, a large part of the phase space here leads to different disordered local minima. However, above the freezing point, the relative phase space volume of ordered structures becomes dominant, regardless of the fact that an immense number of distinct basins exist at the same packing fraction level, corresponding to jammed states with disordered structures.

\begin{figure}[tbhp]
\begin{center}
\includegraphics[width=8.5cm]{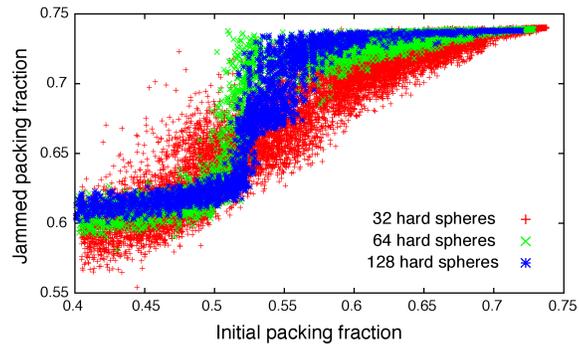}
\end{center}
\caption {Packing fraction of the quenched jammed configurations as a function of their initial packing fraction for the 32, 64 and 128 hard spheres systems.}
\label{fig:Phi_init_jammed}
\end{figure}

The quenching process allows the characterisation of the phase transition without the use of an explicit and somewhat arbitrarily chosen order parameter. In Figure~\ref{fig:Phi_init_jammed} we plot a random subset of our samples with the packing fraction before and after quenching on the horizontal and vertical axes. The phase transition is again visible, showing that configurations just below and above the critical packing fraction are close to jammed states with relatively low and high packing fraction, respectively. 

\end{document}